\documentclass[showpacs,reprint,superscriptaddress, aps, prl, floatfix]{revtex4-1}
\usepackage[utf8]{inputenc}
\usepackage{graphicx, xcolor}
\usepackage{dcolumn}
\usepackage{bm}
\usepackage{amsmath,amsthm,amssymb,bbold}
\usepackage{color}
\usepackage{verbatim}
\usepackage{physics}
\usepackage{nicefrac}
\usepackage{float}
\usepackage{natbib}
\usepackage{wasysym}
\usepackage{amsfonts}
\usepackage{booktabs}
\usepackage{times}
\usepackage{siunitx}
\usepackage[T1]{fontenc}
\usepackage{tikz}

\newcommand{\cwSmall}{\raisebox{-1.2ex}{\includegraphics[scale=.0155]{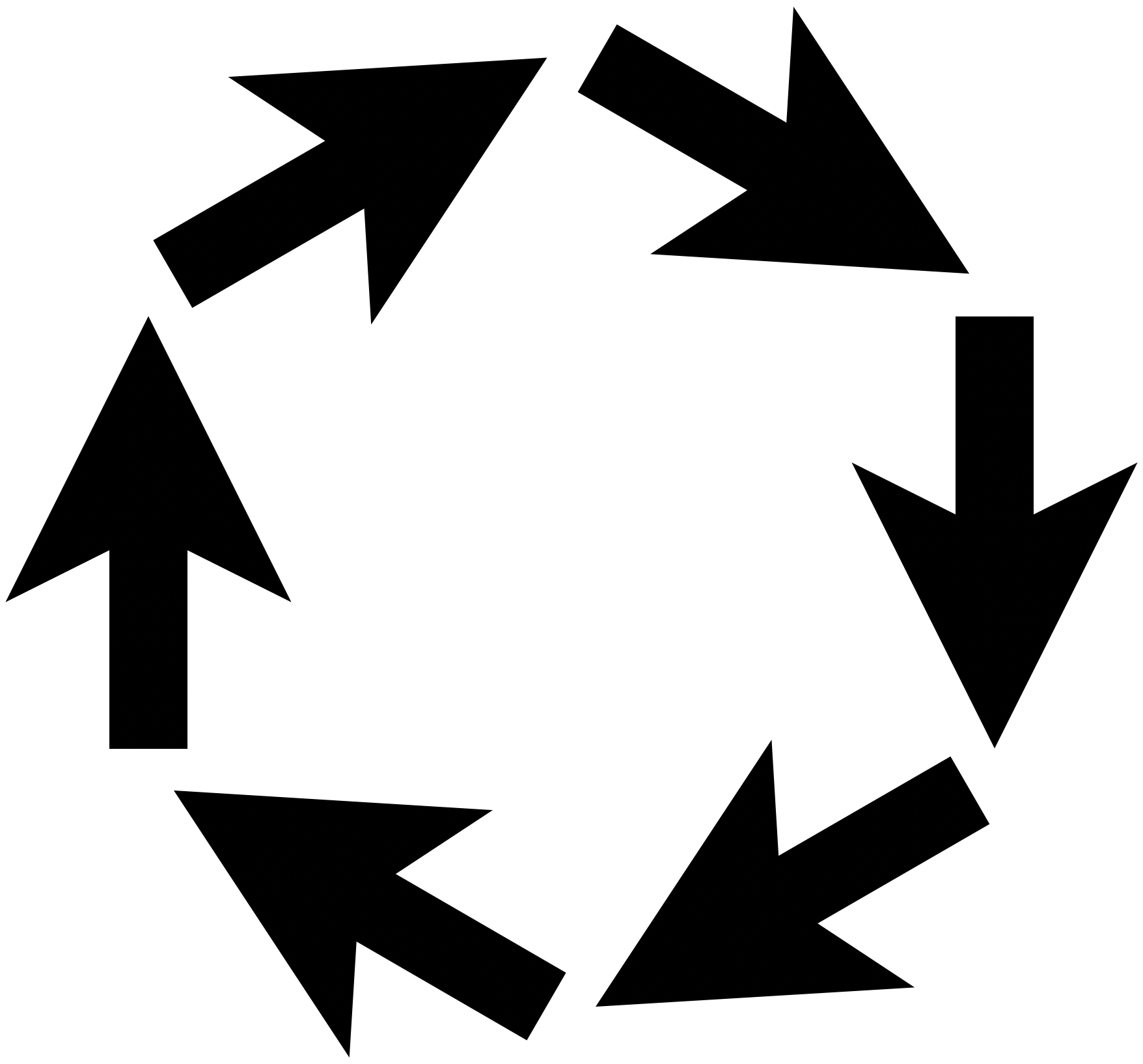}}}

\newcommand{\ccwSmall}{\raisebox{-1.2ex}{\includegraphics[scale=.0155]{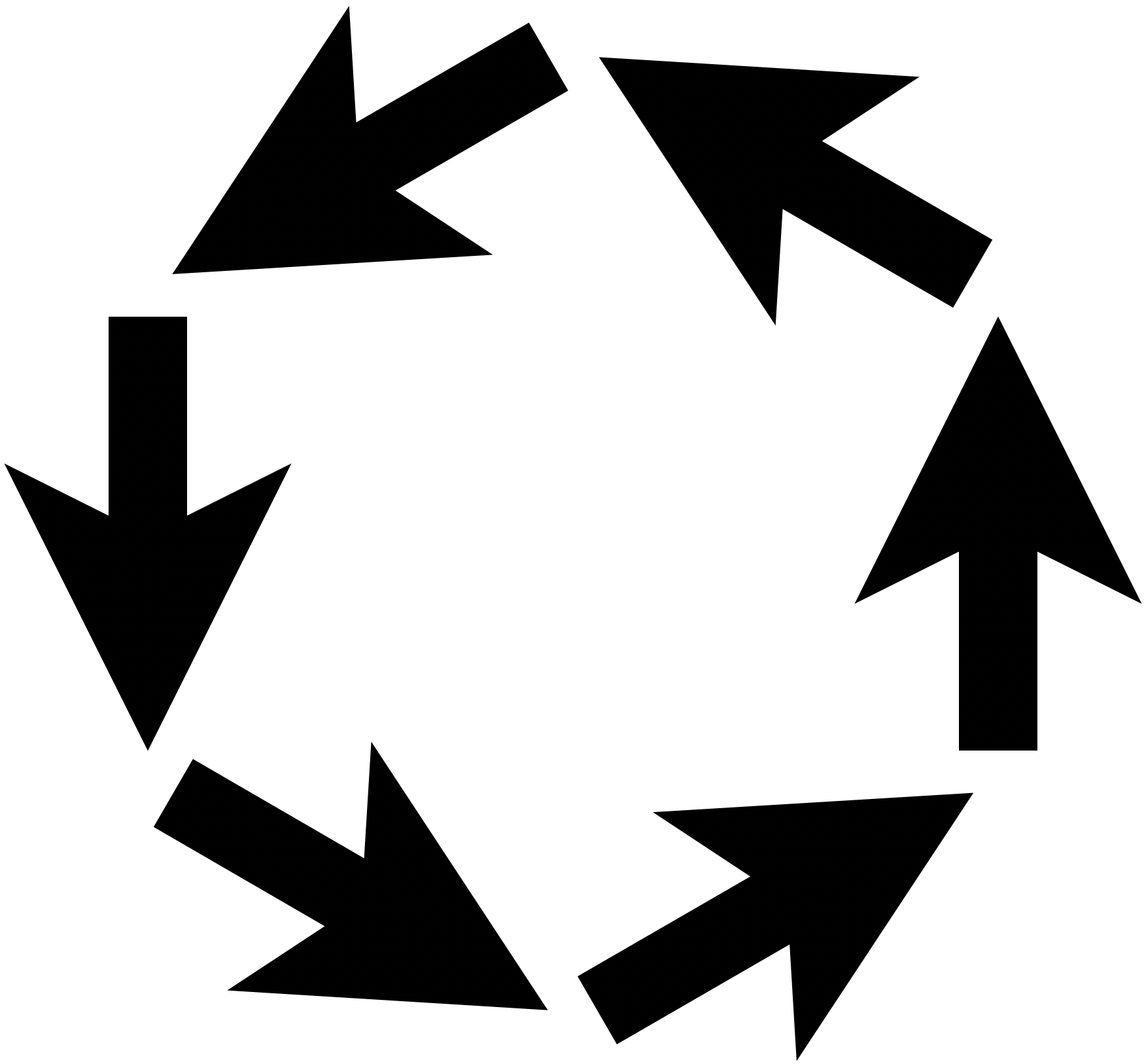}}}

\definecolor{darkGreen}{RGB}{0,110,0}
\definecolor{darkBlue}{RGB}{0,0,130}
\usepackage[colorlinks,citecolor=darkGreen,linkcolor=darkBlue,urlcolor=blue,hyperindex]{hyperref}

\begin{document}
\title{The Emergent Fine Structure Constant of Quantum Spin Ice Is Large}

\author{Salvatore D. Pace}
\affiliation{TCM Group, Cavendish Laboratory, University of Cambridge, Cambridge CB3 0HE, United Kingdom}
\affiliation{Department of Physics, Boston University, Boston, Massachusetts 02215, USA}

\author{Siddhardh C. Morampudi}
\affiliation{Center for Theoretical Physics, Massachusetts Institute of Technology, Cambridge, MA 02139, USA}

\author{Roderich Moessner}
\affiliation{Max-Planck-Institut f\"{u}r Physik komplexer Systeme, 01187 Dresden, Germany}

\author{Chris R. Laumann}
\affiliation{Department of Physics, Boston University, Boston, Massachusetts 02215, USA}

\date{\today}

\begin{abstract}
Condensed matter systems provide alternative `vacua' exhibiting emergent low-energy properties drastically different from those of the standard model.
A case in point is the emergent quantum electrodynamics (QED) in the fractionalized topological magnet known as quantum spin ice, whose magnetic monopoles set it apart from the familiar QED of the world we live in.
Here, we show that the two greatly differ in their fine-structure constant $\alpha$, which parametrizes how strongly matter couples to light: 
$\alpha_{\mathrm{QSI}}$ is more than an order of magnitude greater than $\alpha_{\mathrm{QED}} \approx 1/137$. 
Furthermore, $\alpha_{\mathrm{QSI}}$, the emergent speed of light, and all other parameters of the emergent QED, are tunable by engineering the microscopic Hamiltonian.
We find that $\alpha_{\mathrm{QSI}}$ can be tuned all the way from zero up to what is believed to be the \textit{strongest possible} coupling beyond which QED confines.
In view of the small size of its constrained Hilbert space, this marks out quantum spin ice as an ideal platform for studying exotic quantum field theories and a target for quantum simulation.
The large $\alpha_{\mathrm{QSI}}$ implies that experiments probing candidate condensed-matter realizations of quantum spin ice should expect to observe phenomena arising due to strong interactions.
\end{abstract}

\maketitle

The fine structure constant of QED, $\alpha_{\mathrm{QED}}\approx1/137$, is famously measurable in a semiconductor device~\cite{Klitzing1985}, oblivious to any imperfections of the crystal, and perfectly immutable compared to measurements {\it in vacuo}~\cite{Gabrielse2006}. 
By contrast, a fine structure constant is also known to emerge entirely independently in quantum condensed matter phases whose emergent excitations mimic QED~\cite{Levin2005, Anderson1972}. 
This emergent fine-structure constant has no reason to be as constrained as that in QED and this allows emergent QEDs (eQED) to probe physical regimes which are usually difficult to access either theoretically or experimentally. 

\begin{figure}[t!]
    \centering
    \includegraphics[width=.48\textwidth]{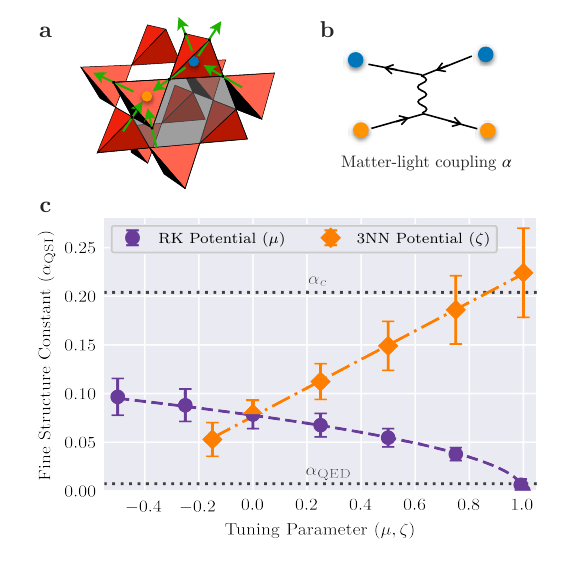}
    \caption{
    \textbf{a)} The pyrochlore lattice of quantum spin ice (QSI) is formed from corner sharing tetrahedra with spin $1/2$s residing at corners. 
    The spins shown give an example of ice-rule violating tetrahedra that correspond to an electric charge-anticharge pair.
    \textbf{b)} The emergent electric charges and photons can interact, just as electrons and photons do in QED, and their interaction strength is given by the emergent fine structure constant, $\alpha_\mathrm{QSI}$.
    \textbf{c)} The value of $\alpha_\mathrm{QSI}$ in the eQED phase of the microscopic QSI Hamiltonian (see Eqn.~\eqref{eqn:QSIp}) shown as a function of $\mu$ (with $\zeta = 0$) and $\zeta$ (with $\mu=0$). 
    Error bars represent the standard deviation of $\alpha_{\mathrm{QSI}}$ among its shape-dependent variations at a fixed ($\mu$,$\zeta$).
    By varying the 3NN potential, $\alpha$ is tunable up to the maximum value $\alpha_c$ (dotted line) beyond which it is conjectured that any compact QED in 3+1D confines~\cite{Cardy1980,Luck1982,Cella1997}.
    }
    \label{fig:eQEDandAlpha}
\end{figure}

Among the various microscopic models which host an eQED~\cite{Levin2005, Wen2001, Motrunich2002, Moessner2003, Hermele2004}, the ones which have received the most attention recently in experiments go under the name of quantum spin ice~\cite{Gingras2014, pan2016measure, sibille2018experimental, gaudet2019quantum, gao2019experimental}. 
The term quantum spin ice (QSI) simultaneously refers to a family of models, as well as a class of rare-earth magnetic materials which approximately realize the theoretical models.
Similar to the prototypical gauge theory of QED which has matter excitations such as electrons and a gauge boson corresponding to the photon, the eQED in QSI is a 3+1D compact $U(1)$ gauge theory and has ``matter'' excitations~\footnote{The terminology for the excitations in QSI differs among communities~\cite{Gingras2014}; we adopt the language used by the gauge theory literature where the spinon is called an electric charge. Our electric charge is referred to as a magnetic monopole in the classical spin ice literature and a spinon in the quantum spin liquid literature. Our magnetic monopole is also sometimes referred to as a vison in the quantum spin ice literature. } --- electric charges (which are the spinons shown in Fig.~\ref{fig:eQEDandAlpha}a) and magnetic monopoles --- and an (emergent) photon~\cite{Hermele2004, kogut1979introduction}.
These emergent photons and spinons have been established by various theoretical and numerical studies~\cite{Banerjee2008, Shannon2012, Benton2012, Kato2015}, with the magnetic monopoles being the focus of recent studies~\cite{Szabo2019a,PhysRevB.102.125113}.
Understanding the properties of the eQED necessitates not just identification of the low-energy emergent excitations, but also measuring the various couplings of the eQED such as the speed of light $c_{\mathrm{QSI}}$. 
These can be drastically different from those of usual QED, giving access to unusual regimes and phenomenology typically inaccessible in our world. For example, estimates of $c_{\mathrm{QSI}}$ are quite small~\cite{Hermele2004, Benton2012, sibille2018experimental}. 
This means that experiments can probe phenomena ranging from the non-relativistic to the ultra-relativistic, where the electric charges move faster than the speed of light and emit Cerenkov radiation.

However, there is currently no estimate of the electric charge $e_{\mathrm{QSI}}$ and hence fine-structure constant $\alpha_{\mathrm{QSI}} \equiv e^2_{\mathrm{QSI}}/\hbar c_{\mathrm{QSI}}$ (in fact, in any microscopic model with an eQED).
This dimensionless quantity characterizes how strongly the spinons (which are the electric charges of the theory) interact with the emergent photon (see Fig.\ref{fig:eQEDandAlpha}b).
In usual QED, the small value of $\alpha_{\mathrm{QED}}$ justifies a perturbative treatment, while also making some processes like photon-photon scattering very difficult to observe.
Determining the value of $\alpha_{\mathrm{QSI}}$ would allow us to guide theoretical treatments of its eQED and also potentially place the eQED in a different regime to QED.

Here, we determine the fine-structure constant $\alpha_{\mathrm{QSI}}$ in the eQED of QSI. 
Besides being an order of magnitude larger than $\alpha_{\mathrm{QED}}$, it is \emph{tunable} over the complete theoretical range by adding local interactions to the microscopic Hamiltonian. 
This also constitutes a clear example where modifying the microscopic details of a theory changes the emergent couplings of the low-energy theory in a straightforward manner.
Our main results are displayed in Fig.~\ref{fig:eQEDandAlpha}c and in Table~\ref{table:summary}. 
From a methodological perspective, the framework we have developed using large-scale exact diagonalization (ED) techniques in constrained spaces may be of additional interest in determining the low-energy properties of other microscopic models with exotic emergent theories.

\emph{Microscopics---}
Spin ice is modeled by spin-$1/2$ particles residing on the corners of the tetrahedra of the pyrochlore lattice, shown in Fig.~\ref{fig:eQEDandAlpha}a~\cite{Gingras2014}.
Each spin is restricted to point either toward or away from the centers of the two adjacent tetrahedra.
With this restriction, the classical ground state follows a simple rule~\cite{Bramwell2001}: each tetrahedron has two spins pointing in and two pointing out. 
This ``2-in 2-out'' local constraint is called the ice rule, named after a similar constraint in water ice~\cite{Anderson1956}. 
Classical spin ice is well understood in terms of fractionalized spins forming an emergent classical electromagnetism, with the ice rule playing the role of Gauss's law. 
Local violations of the ice rule then correspond to spinons and antispinons~\cite{Castelnovo2012}, which we refer to as electric charges and anticharges.
At low temperatures, quantum fluctuations allow tunneling between classical configurations satisfying the ice rule, giving rise to an eQED~\cite{Hermele2004, Banerjee2008, Shannon2012, Huang2018, Ross2011, Kato2015, Benton2012, Savary2012, Lee2012}.
In addition to the electric charges, there are now magnetic monopoles as well as photons corresponding to coherent ring-exchange processes within the ice manifold.

The microscopic Hamiltonian to describe QSI materials was derived, and studied in considerable detail, in the context of the rare earth pyrochlore materials~\cite{Gingras2014, Ross2011, Rau2019}. 
For the present purposes, it is sufficient to consider a simplified model given by the canonical QSI Hamiltonian which consists of two parts~\cite{Hermele2004}: 
a `classical' term enforcing the ice rules, which determines the cost of an electric charge; 
and a `quantum' resonance term, also known as a loop flip or ring exchange term, $W_{\hexagon}$, which coherently flips a sequence of six spins arranged head to tail around a hexagon \raisebox{4pt}{\cwSmall\hspace{2pt}$\to$ \ccwSmall},
\begin{equation}\label{eqn:QSIeff}
    {H_{\mathrm{eff}}}=
    J_{z z}\sum_{\langle i,j\rangle} S_{i}^{z}S_{j}^{z}
    -g\sum_{\hexagon}\left(W_{\hexagon} + W_{\hexagon}^{\dagger}\right).
\end{equation}
The first sum runs over all bonds of the pyrochlore lattice and the second over all of its hexagonal plaquettes. 
A hexagonal plaquettes on which $W_{\hexagon}$ acts is shaded in Fig.~\ref{fig:eQEDandAlpha}a. 
This Hamiltonian describes the standard low-energy dynamics of geometrically frustrated systems capturing phenomena ranging from high-temperature superconductivity to frustrated magnetism~\cite{Moessner2011}, and can be obtained as a low-energy effective theory of the general microscopic quantum spin ice model~\cite{Hermele2004,Savary2012}
Furthermore, it can be formally rewritten as a compact $U(1)$ lattice gauge theory~\cite{Hermele2004, Shannon2012}, with $W_{\hexagon}$ the smallest possible Wilson loop.

To effect the above-mentioned tuning, we additionally consider a pair of simple perturbations to $H_{\mathrm{eff}}$:
\begin{equation}\label{eqn:QSIp}
    {H_{\mathrm{p}}}=
\zeta \hspace{-6pt}\sum_{\langle\langle\langle i,j\rangle\rangle\rangle} \hspace{-6pt}S_{i}^{z}S_{j}^{z} +  \mu \sum_{\hexagon}\left( W_{\hexagon}^{\dagger}W_{\hexagon} + W_{\hexagon}W_{\hexagon}^{\dagger} \right).
\end{equation}
The first summation over $\langle\langle\langle i,j\rangle\rangle\rangle$ runs over the third-nearest neighbors (3NN), which are pairs of spins across from each other on a hexagonal plaquette. 
This two-body Ising term generically exists in material realizations~\cite{Ross2011} and can be engineered in many current quantum simulators~\cite{King2020,Bernien:2017to,Pagano:2020wp}.
It prefers spins across from each other to be (anti)parallel (depending on the sign of $\zeta$), hence affecting the number of flippable hexagons (\raisebox{4pt}{\cwSmall}). 
The second term is a Rokhsar-Kivelson (RK) potential, which directly counts the number of flippable hexagons and, as a six body term, is less easy to control experimentally.
However, the ground state is exactly solvable at the RK point~\cite{Rokhsar1988} ($\zeta = 0 $and $\mu = 1$) which allows us to validate our numerics by comparing to previous analytic and numerical studies~\cite{Moessner2003, Shannon2012, Benton2012}. 
We note that tuning either of these perturbations to be sufficiently strong causes the system to transition out of the deconfined QED phase, which we find persists for $-0.5 \lesssim \mu \leq 1$ at $\zeta=0$~\cite{Shannon2012}, and for $-0.2 \lesssim \zeta \lesssim 1$ at $\mu=0$ (see supplemental materials).

\emph{Macroscopic eQED---}%
The low-energy theory of eQED is the familiar Maxwell Hamiltonian
\begin{equation}\label{eqn:Maxwell}
    H_{\mathrm{Maxwell}} = \frac{1}{8\pi}\int d^{3}\bm{x} \left( \left|\bm{E}\right|^{2} +
    c_{\mathrm{QSI}}^2\left|\bm{B}\right|^{2} \right),
\end{equation}
where $\bm{B}=\operatorname{curl}\bm{A}$, and $\bm{E}$ and $\bm{A}$ are the canonically conjugate electric field and vector potential operators, respectively. 
Throughout this manuscript, we use units such that the emergent Coulomb energy between two electric charges (magnetic monopoles) is $e_{\mathrm{QSI}}^2/r$ ($m_{\mathrm{QSI}}^2/r$).
We fit the low-energy spectra of Eqn.~\eqref{eqn:QSIeff} in the constrained Hilbert space obeying the classical ice rules, using results from Eqn.~\eqref{eqn:Maxwell} to extract $e_{\mathrm{QSI}}$ and $c_{\mathrm{QSI}}$. See the supplemental materials for a detailed account of the ED techniques used to access the spectra of systems with up to 96 spins.

Since electric charges cannot be excited in the constrained Hilbert space, it may appear that $e_{\mathrm{QSI}}$ cannot be probed. 
However, it is possible to have electric field lines looping through the periodic boundaries without violating the ice rules~\cite{Hermele2004,Shannon2012}. 
As a gedanken experiment, an elementary unit of the electric field can be created by first exciting an electric charge-anticharge pair, moving the electric charge around the lattice through a periodic boundary, and then annihilating it with the electric anticharge.
This leaves behind an elementary unit of electric flux passing through the boundary. 
As the dynamics of the QSI Hamiltonian preserve the ice rule locally, the Hilbert space decomposes into electric topological sectors $\bm{\phi} = (\phi_{1}, \phi_{2}, \phi_{3})\in\mathbb{Z}^{3}$, where $\phi_{i}$ gives the number of elementary units of electric flux through the $i^{\text{th}}$ direction.

\begin{figure}[t!]
    \centering
    \includegraphics[width=.48\textwidth]{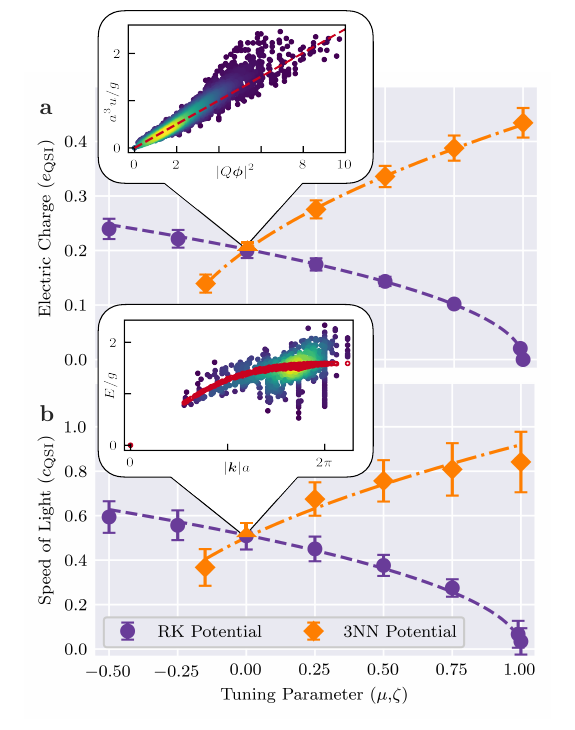}
    \caption{
    \textbf{a)} The emergent electric charge $e_{\mathrm{QSI}}$ as a function of RK ($\mu$,$\zeta = 0$) and 3NN ($\mu=0$, $\zeta$) potential.
    A representative scatter plot of this data is shown in the inset (data corresponding to $\mu = \zeta = 0$) with associated fit (red line). 
    The dashed lines are fits giving $e_{\mathrm{QSI}} = 0.20\sqrt{ag(1-\mu)}$ at $\zeta = 0$ and  $e_{\mathrm{QSI}} = 0.38\sqrt{ag(0.28+\zeta)}$ at $\mu = 0$. 
    We note that the former dependence is predicted near the RK point at $\mu = 1$~\cite{Moessner2003, Shannon2012}, while the latter is a guide to the eye.
    \textbf{b)} The emergent speed of light $c_{\mathrm{QSI}}$ as a function of RK and 3NN potential.
    Representative scatter plot of this dispersion is shown in the inset (at $\mu = \zeta = 0$) with associated fit (red line). 
    Dashed lines are fits giving $c_{\mathrm{QSI}} = 0.51ag\sqrt{1-\mu}/\hbar$ and $c_{\mathrm{QSI}} = 0.78ag\sqrt{0.41 + \zeta}/\hbar$ along the $\zeta = 0$ and $\mu = 0$ axes, respectively.
    Again, we note that the dependence of $c$ on $\mu$ near the RK point is consistent with previous results~\cite{Moessner2003, Benton2012}.
    The error bars in both panels represent the standard deviation of $e_{\mathrm{QSI}}$ and $c_{\mathrm{QSI}}$ among its shape-dependent variations at a fixed ($\mu$,$\zeta$).
    Furthermore, in both insets, scatter points are brighter the denser their neighboring data points are.
    }
    \label{fig:e2_and_c_data}
\end{figure}

The electric field created by this procedure is uniform when the lattice is coarse-grained. 
By computing the ground state energy in each electric topological sector, we can thus extract the value of $e_{\mathrm{QSI}}$. 
As shown in the supplemental material, $\bm{E}$ can be found using Gauss's law which then gives an expression for the electric field energy density
\begin{equation}\label{eqn:uE}
    u =  e_{\mathrm{QSI}}^{2}\frac{2\pi|Q\bm{\phi}|^{2}}{a^{4}},
\end{equation}
where $a$ is the lattice constant of the face-centered cubic lattice underlying the pyrochlore lattice and $Q$ is a dimensionless $3\times 3$ matrix characterizing the shape of the periodic volume. 
The inset of Fig.~\ref{fig:e2_and_c_data}a shows the fit of Eqn.~\eqref{eqn:uE} to the $u$ ED data at $\mu = \zeta = 0$, yielding $e_{\mathrm{QSI}} = 0.20(1)\sqrt{ag}$.
The ED data is obtained across a range of finite-size samples (up $N=96$ spins and $180$ different shapes). 
The spread of the data about the fit, and the corresponding variation in $e_{\mathrm{QSI}}$, comes from the variations in the measurement for different lattice shapes occurring due to the limited sizes accessible with ED. 

Fig.~\ref{fig:e2_and_c_data}a shows $e_{\mathrm{QSI}}$ measured at different values of $\zeta$ and $\mu$ in Eqn.~\eqref{eqn:QSIp} along the $\mu = 0$ and $\zeta = 0$ axes, respectively. 
As $\zeta$ becomes increasingly positive and $\mu$ increasingly negative, $e_{\mathrm{QSI}}$ increases. This has a simple interpretation. 
Both of these perturbations increase the microscopic energy for spins across hexagonal plaquettes to be parallel, which in terms of the eQED correspond to states with local electric flux in the direction of the parallel spins. 
This increases the energy of the sectors with global electric flux, producing a larger $e_{\mathrm{QSI}}$.

We measure $c_{\mathrm{QSI}}$ using the ground state dispersion of Eqn.~\eqref{eqn:QSIeff} translated into the first Brillouin zone.
At small momenta, one of the photon's key characteristics is its relativistic dispersion $\omega(\bm{k}) = c_{\mathrm{QSI}}|\bm{k}|$. 
The ED data used to extract the fit is obtained across the same range of samples as in the measurement for $e_{\mathrm{QSI}}$.
We obtain the value of $c_{\mathrm{QSI}}$ by using the Gaussian approximation to the photon dispersion on the pyrochlore (see supplemental material for derivation):~\cite{Benton2012}
\begin{equation}\label{eqn:dispersion}
\omega(\bm{k})  = \sqrt{\frac{c^{2}_{\mathrm{QSI}}}{a^{2}}\lambda(\bm{k}) + M\lambda^{2}(\bm{k})},
\end{equation}
where $c_{\mathrm{QSI}}$ and $M$ are fitting parameters and $\lambda(\bm{k})=12-4\sum_{i > j}\cos \left(k_{i} a/2\right) \cos \left(k_{j} a/2\right)$.
The inset of Fig.~\ref{fig:e2_and_c_data}b shows the momentum dependence of the ground state energy at $\mu = \zeta = 0$, which upon fitting Eqn.~\eqref{eqn:dispersion} gives $c_{\mathrm{QSI}} = 0.51(6)ag/\hbar$.
In addition to variation of $c_{\mathrm{QSI}}$ coming from lattice shape dependence, there may be spread from the fit due to magnetic monopole states at higher momenta that the Gaussian photon dispersion does not capture~\cite{Huang2018}; in particular, we exclude $|k|a > \pi$ from the fit, where clear irregularities are visible. 
We note that the fit value is similar to a previous numerical measurement $c_{\mathrm{QSI}} = 0.6(1)ag/\hbar$~\cite{Shannon2012} and analytical estimate $c = 0.41ag / \hbar$~\cite{Kwasigroch2017} using semi-classical techniques. 

Using the ED spectra along the $\mu$ and $\zeta$ axes, Fig.~\ref{fig:e2_and_c_data}b shows that like $e_{\mathrm{QSI}}$, $c_{\mathrm{QSI}}$ is indeed also tunable. 
We see a similar trend as previously: $c_{\mathrm{QSI}}$ increases as states with a greater number of flippable hexagons become energetically favored. 
This can be understood qualitatively by noting that the photons are collective motions of fluctuating electric field loops~\cite{Levin2005}. 
Since a hexagon has to be flippable to support local electric field loop fluctuations, the photon can propagate to flippable hexagons more rapidly than unflippable ones. 
At long wavelengths, this corresponds to an increase of the speed of light with increased density of flippable hexagons.

\emph{Fine structure constant---} In our units, the fine structure constant is given by $\alpha = e^{2}/ \hbar c$. From our measurements of $e_{\mathrm{QSI}}$ and $c_{\mathrm{QSI}}$, upon taking their quotient to find $\alpha_{\mathrm{QSI}}$ the dimensionful constants $a$ and $g$ crucially cancel. Fig.~\ref{fig:eQEDandAlpha}c shows $\alpha_{\mathrm{QSI}}$ as a function of $\zeta$ and $\mu$ along the $\mu = 0$ and $\zeta = 0$ axes. Varying $\mu$, we see that $\alpha_{\mathrm{QSI}}$ is tunable ranging from exactly zero at the RK point all the way to $0.1$ at $\mu = -0.5$, beyond which the system undergoes a first order transition into an ordered state~\cite{Shannon2012}. 
Along the $\mu = 0$ axis, $\alpha_{\mathrm{QSI}}$ is $0.06$ at $\zeta = -0.15$ and increases to $0.2$ at $\zeta = 1$. 
At $\zeta \approx 1$, the Hamiltonian undergoes a phase transition into a finite momentum phase, suggesting the development of long-range magnetic order and  confinement of the eQED (see supplemental material). 
It is remarkable to note that the value $\alpha_{\mathrm{QSI}}$ takes at $\zeta = 1$ corresponds to $\alpha_{c} \approx 0.2$ at which pure lattice QED on the cubic lattice is known to confine~\cite{Jersak1983}. 
Indeed, $\alpha_{c} \approx 0.2$ has been argued to be the limit of stability of the deconfined phase in general~\cite{Cardy1980,Luck1982,Cella1997}. 
Thus, we find that we can tune $\alpha_{\mathrm{QSI}}$ over the entire range of fine structure constants allowed by a deconfined QED: $0 \leq \alpha \leq 0.2$.

\begin{table}[t!]
\begin{ruledtabular}
{\renewcommand{\arraystretch}{1.5}
\begin{tabular}{ccc}
 &  Candidate QSI Material & Vacuum QED \\
\hline

$\alpha$ & $1/10$ & $1/137$ \\

$c$  & $1$ \si{\metre\per\second} & $3.0 \times 10^{8}$ \si{\metre\per\second}\\

$e$ & $10^{-4}$ $\sqrt{\si{\electronvolt\nano\metre}}$ & $1.2$ $\sqrt{\si{\electronvolt\nano\metre}}$\\

$m$ & $10^{-3}$ $\sqrt{\si{\electronvolt\nano\metre}}$ & $82.2$ $\sqrt{\si{\electronvolt\nano\metre}}$ 

\end{tabular}}
\end{ruledtabular}
\caption{%
Numerical values of the fine structure constant, $\alpha = e^2 / \hbar c$, the speed of light, $c$, the elementary electric charge $e$, and the elementary magnetic charge from Dirac quantization, $m = e / 2 \alpha$. 
In our units, the electric (magnetic) charge squared corresponds to the energy between two electric charges (magnetic monopoles) held one nanometer apart. 
The second column uses characteristic scales obtained from the pyrochlore oxides, corresponding to $\mu = \zeta = 0$, $a = 10$\si{\angstrom}, and $g \sim 1$\si{\micro\electronvolt}. 
We stress that the dimensionful values of $a$ and $g$ do not affect $\alpha_\mathrm{QSI}$. 
The corresponding values in the vacuum QED of our universe are shown in the third column.
}
\label{table:summary}
\end{table}

The dimensionful quantities $e_{\mathrm{QSI}}$ and $c_{\mathrm{QSI}}$ we have calculated depend on the lattice parameters $a$ and $g$. 
There are a large variety of rare-earth pyrochlore oxides that are QSI candidates, such as Tb$_{2}$Ti$_{2}$O$_{7}$, Yb$_{2}$Ti$_{2}$O$_{7}$, Pr$_{2}$Sn$_{2}$O$_{7}$, and Pr$_{2}$Zr$_{2}$O$_{7}$~\cite{Gingras2014,Rau2019, Gardnner2010}. 
The lattice constant in these materials are approximately $a \approx 10$\si{\angstrom}~\cite{Gardnner2010} and typical energy values of a candidate QSI material correspond to $g \approx 1$\si{\micro\electronvolt}~\cite{Savary2012, Ross2011}. 
Using these values, we can estimate $e_{\mathrm{QSI}}$ and $c_{\mathrm{QSI}}$, which are shown in table \ref{table:summary} along with the corresponding values in vacuum QED. 
This highlights the exotic nature of the eQED in QSI: the emergent photon travels \emph{a hundred million times slower} than the speed of light and the emergent fine structure constant is \emph{ten times larger} than its vacuum QED counterpart. 
The largeness of $\alpha_{\mathrm{QSI}}$ implies substantial interactions between spinons and emergent photons in QSI, consistent with deviations from non-interacting theory expectations for the dynamic structure factor observed in quantum Monte Carlo at finite temperature~\cite{Huang2018}

The experimental effort to establish that these candidate materials realize the deconfined eQED phase at low temperature have largely been focused on finding evidence for the existence of a linearly dispersing transverse photon and fractionalized gapped spinons -- the non-interacting structure of the emergent particles.
However, the size of $\alpha_{\mathrm{QSI}}$ suggest that distinctive experimental signatures may actually follow from the interaction effects between the particles.
For example, due to $\alpha_{\mathrm{QSI}}$, we expect the dynamic structure factor observed in neutron scattering to exhibit the presence of well-defined spinon-antispinon `Rydberg' bound states, a strong Sommerfeld enhancement of the pair-production continuum at small momenta, and a strong diffusive suppression of the continuum at large momenta due to emergent Cerenkov radiation within the sample~\cite{Morampudi2020}.
Observation of such effects would thus constitute strong evidence for the eQED phase in these materials. 
The values of the constants determined here are inputs for quantitative comparison between theory and such experiments.

Finally, we note that our results makes QSI a particularly attractive target for noisy intermediate-scale quantum simulations~\cite{Preskill2018} of strongly coupled, deconfined QED in other  experimental platforms. 
The microscopic construction requires only one two-level qubit per lattice link coupled by two-body local interactions; there have accordingly been detailed engineering proposals in, for example, ultracold Rydberg atoms~\cite{Glaetzle2014}, and demonstrations of closely related 2D ice in superconducting annealers~\cite{King2020}.
The Schwinger model of (1+1)D QED has in fact been realized in multiple quantum simulators recently~\cite{Martinez2016,Yang2020}; however, it only exists in the confined phase.
Our results show that the simple 3NN term $(\zeta)$ provides a direct tuning parameter for the emergent fine structure constant over a broad range to the strongest available coupling, allowing the controlled experimental investigation of strong coupling QED phenomena in (3+1)D.
By varying $\zeta$ in space or time, this also gives a natural setting for studying the consequences of a space-time dependent fine structure constant -- which contrasts with the usual QED where a large amount of effort concludes no such variation exists~\cite{Wilczynska2020}.
By varying the temperature and the corresponding density of emergent matter excitations, this further provides a platform for studying the behavior of strongly coupled plasma containing both electric charges and magnetic monopoles. 

Originally introduced by Sommerfeld~\cite{Sommerfeld1916} to describe the fine structure of the spectral lines in Hydrogen, the smallness of the fine structure constant $\alpha \sim 1/137$ has evolved into one of the great mysteries of our universe. 
Its smallness enables the description of physical law in terms of weakly coupled matter and light, even as the largeness of $1/\alpha \sim 137$ determines the maximum stable atomic numbers of the periodic table and thus the richness of chemistry. 
However, despite almost a century of effort, there is no microscopic grand unified theory which predicts this fundamental parameter of our universe.
By studying the emergent phenomena provided by the strongly coupled eQED of spin ice, perhaps new light can be shed on this fundamental enigma.

\textbf{Acknowledgements}
The authors are grateful to Subhro Bhattacharjee, Anushya Chandran, Bert Halperin, and Frank Wilczek for discussions. 
We thank Boston University’s Research Computing Services for their computational resources.
S.D.P. acknowledges support from Boston University’s Undergraduate Research Opportunities Program and from The Winston Churchill Foundation of the United States through the Churchill Scholarship.
S.M. acknowledges funding from the Tsung-Dao Lee Institute.
This work was in part supported by the Deutsche Forschungsgemeinschaft  under grants SFB 1143 (project-id 247310070) and the cluster of excellence ct.qmat (EXC 2147, project-id 390858490).  
C.R.L. acknowledges support from the NSF through grant PHY-1752727. 
The authors wish to acknowledge the generous hospitality of the Galileo Galilei Institute for Theoretical Physics where this work was initiated and from the Aspen Center for Physics, which is supported by NSF grant PHY-1607611.

\bibliographystyle{apsrev4-1}
\bibliography{references}

\begin{thebibliography}{50}%
\makeatletter
\providecommand \@ifxundefined [1]{%
 \@ifx{#1\undefined}
}%
\providecommand \@ifnum [1]{%
 \ifnum #1\expandafter \@firstoftwo
 \else \expandafter \@secondoftwo
 \fi
}%
\providecommand \@ifx [1]{%
 \ifx #1\expandafter \@firstoftwo
 \else \expandafter \@secondoftwo
 \fi
}%
\providecommand \natexlab [1]{#1}%
\providecommand \enquote  [1]{``#1''}%
\providecommand \bibnamefont  [1]{#1}%
\providecommand \bibfnamefont [1]{#1}%
\providecommand \citenamefont [1]{#1}%
\providecommand \href@noop [0]{\@secondoftwo}%
\providecommand \href [0]{\begingroup \@sanitize@url \@href}%
\providecommand \@href[1]{\@@startlink{#1}\@@href}%
\providecommand \@@href[1]{\endgroup#1\@@endlink}%
\providecommand \@sanitize@url [0]{\catcode `\\12\catcode `\$12\catcode
  `\&12\catcode `\#12\catcode `\^12\catcode `\_12\catcode `\%12\relax}%
\providecommand \@@startlink[1]{}%
\providecommand \@@endlink[0]{}%
\providecommand \url  [0]{\begingroup\@sanitize@url \@url }%
\providecommand \@url [1]{\endgroup\@href {#1}{\urlprefix }}%
\providecommand \urlprefix  [0]{URL }%
\providecommand \Eprint [0]{\href }%
\providecommand \doibase [0]{http://dx.doi.org/}%
\providecommand \selectlanguage [0]{\@gobble}%
\providecommand \bibinfo  [0]{\@secondoftwo}%
\providecommand \bibfield  [0]{\@secondoftwo}%
\providecommand \translation [1]{[#1]}%
\providecommand \BibitemOpen [0]{}%
\providecommand \bibitemStop [0]{}%
\providecommand \bibitemNoStop [0]{.\EOS\space}%
\providecommand \EOS [0]{\spacefactor3000\relax}%
\providecommand \BibitemShut  [1]{\csname bibitem#1\endcsname}%
\let\auto@bib@innerbib\@empty
%</preamble>
\bibitem [{\citenamefont {von Klitzing}(1993)}]{Klitzing1985}%
  \BibitemOpen
  \bibfield  {author} {\bibinfo {author} {\bibfnamefont {K.}~\bibnamefont {von
  Klitzing}},\ }in\ \href@noop {} {\emph {\bibinfo {booktitle} {Nobel Lectures
  in Physics 1981 -- 1990}}},\ \bibinfo {editor} {edited by\ \bibinfo {editor}
  {\bibfnamefont {G.}~\bibnamefont {Ekspong}}}\ (\bibinfo  {publisher} {World
  Scientific Publishing Co.},\ \bibinfo {address} {Singapore},\ \bibinfo {year}
  {1993})\ pp.\ \bibinfo {pages} {309--346}\BibitemShut {NoStop}%
\bibitem [{\citenamefont {Gabrielse}\ \emph {et~al.}(2006)\citenamefont
  {Gabrielse}, \citenamefont {Hanneke}, \citenamefont {Kinoshita},
  \citenamefont {Nio},\ and\ \citenamefont {Odom}}]{Gabrielse2006}%
  \BibitemOpen
  \bibfield  {author} {\bibinfo {author} {\bibfnamefont {G.}~\bibnamefont
  {Gabrielse}}, \bibinfo {author} {\bibfnamefont {D.}~\bibnamefont {Hanneke}},
  \bibinfo {author} {\bibfnamefont {T.}~\bibnamefont {Kinoshita}}, \bibinfo
  {author} {\bibfnamefont {M.}~\bibnamefont {Nio}}, \ and\ \bibinfo {author}
  {\bibfnamefont {B.}~\bibnamefont {Odom}},\ }\href {\doibase
  10.1103/PhysRevLett.97.030802} {\bibfield  {journal} {\bibinfo  {journal}
  {Phys. Rev. Lett.}\ }\textbf {\bibinfo {volume} {97}},\ \bibinfo {pages}
  {030802} (\bibinfo {year} {2006})}\BibitemShut {NoStop}%
\bibitem [{\citenamefont {Levin}\ and\ \citenamefont {Wen}(2005)}]{Levin2005}%
  \BibitemOpen
  \bibfield  {author} {\bibinfo {author} {\bibfnamefont {M.}~\bibnamefont
  {Levin}}\ and\ \bibinfo {author} {\bibfnamefont {X.-G.}\ \bibnamefont
  {Wen}},\ }\href
  {https://journals.aps.org/rmp/abstract/10.1103/RevModPhys.77.871} {\bibfield
  {journal} {\bibinfo  {journal} {Rev. Mod. Phys.}\ }\textbf {\bibinfo {volume}
  {77}},\ \bibinfo {pages} {871} (\bibinfo {year} {2005})}\BibitemShut
  {NoStop}%
\bibitem [{\citenamefont {Anderson}(1972)}]{Anderson1972}%
  \BibitemOpen
  \bibfield  {author} {\bibinfo {author} {\bibfnamefont {P.~W.}\ \bibnamefont
  {Anderson}},\ }\href {https://science.sciencemag.org/content/177/4047/393}
  {\bibfield  {journal} {\bibinfo  {journal} {Science}\ }\textbf {\bibinfo
  {volume} {177}},\ \bibinfo {pages} {393} (\bibinfo {year}
  {1972})}\BibitemShut {NoStop}%
\bibitem [{\citenamefont {Cardy}(1980)}]{Cardy1980}%
  \BibitemOpen
  \bibfield  {author} {\bibinfo {author} {\bibfnamefont {J.~L.}\ \bibnamefont
  {Cardy}},\ }\href {\doibase 10.1016/0550-3213(80)90417-4} {\bibfield
  {journal} {\bibinfo  {journal} {Nucl. Phys. B}\ }\textbf {\bibinfo {volume}
  {170}},\ \bibinfo {pages} {369} (\bibinfo {year} {1980})}\BibitemShut
  {NoStop}%
\bibitem [{\citenamefont {Luck}(1982)}]{Luck1982}%
  \BibitemOpen
  \bibfield  {author} {\bibinfo {author} {\bibfnamefont {J.}~\bibnamefont
  {Luck}},\ }\href {\doibase 10.1016/0550-3213(82)90253-x} {\bibfield
  {journal} {\bibinfo  {journal} {Nucl. Phys. B}\ }\textbf {\bibinfo {volume}
  {210}},\ \bibinfo {pages} {111} (\bibinfo {year} {1982})}\BibitemShut
  {NoStop}%
\bibitem [{\citenamefont {Cella}\ \emph {et~al.}(1997)\citenamefont {Cella},
  \citenamefont {Heller}, \citenamefont {Mitrjushkin},\ and\ \citenamefont
  {Vicer{\'e}}}]{Cella1997}%
  \BibitemOpen
  \bibfield  {author} {\bibinfo {author} {\bibfnamefont {G.}~\bibnamefont
  {Cella}}, \bibinfo {author} {\bibfnamefont {U.~M.}\ \bibnamefont {Heller}},
  \bibinfo {author} {\bibfnamefont {V.~K.}\ \bibnamefont {Mitrjushkin}}, \ and\
  \bibinfo {author} {\bibfnamefont {A.}~\bibnamefont {Vicer{\'e}}},\ }\href
  {\doibase 10.1103/physrevd.56.3896} {\bibfield  {journal} {\bibinfo
  {journal} {Phys. Rev. D}\ }\textbf {\bibinfo {volume} {56}},\ \bibinfo
  {pages} {3896} (\bibinfo {year} {1997})}\BibitemShut {NoStop}%
\bibitem [{\citenamefont {Wen}(2001)}]{Wen2001}%
  \BibitemOpen
  \bibfield  {author} {\bibinfo {author} {\bibfnamefont {X.-G.}\ \bibnamefont
  {Wen}},\ }\href {\doibase 10.1103/PhysRevLett.88.011602} {\bibfield
  {journal} {\bibinfo  {journal} {Phys. Rev. Lett.}\ }\textbf {\bibinfo
  {volume} {88}},\ \bibinfo {pages} {011602} (\bibinfo {year}
  {2001})}\BibitemShut {NoStop}%
\bibitem [{\citenamefont {Motrunich}\ and\ \citenamefont
  {Senthil}(2002)}]{Motrunich2002}%
  \BibitemOpen
  \bibfield  {author} {\bibinfo {author} {\bibfnamefont {O.~I.}\ \bibnamefont
  {Motrunich}}\ and\ \bibinfo {author} {\bibfnamefont {T.}~\bibnamefont
  {Senthil}},\ }\href {\doibase 10.1103/PhysRevLett.89.277004} {\bibfield
  {journal} {\bibinfo  {journal} {Phys. Rev. Lett.}\ }\textbf {\bibinfo
  {volume} {89}},\ \bibinfo {pages} {277004} (\bibinfo {year}
  {2002})}\BibitemShut {NoStop}%
\bibitem [{\citenamefont {Moessner}\ and\ \citenamefont
  {Sondhi}(2003)}]{Moessner2003}%
  \BibitemOpen
  \bibfield  {author} {\bibinfo {author} {\bibfnamefont {R.}~\bibnamefont
  {Moessner}}\ and\ \bibinfo {author} {\bibfnamefont {S.~L.}\ \bibnamefont
  {Sondhi}},\ }\href {\doibase 10.1103/PhysRevB.68.184512} {\bibfield
  {journal} {\bibinfo  {journal} {Phys. Rev. B}\ }\textbf {\bibinfo {volume}
  {68}},\ \bibinfo {pages} {184512} (\bibinfo {year} {2003})}\BibitemShut
  {NoStop}%
\bibitem [{\citenamefont {Hermele}\ \emph {et~al.}(2004)\citenamefont
  {Hermele}, \citenamefont {Fisher},\ and\ \citenamefont
  {Balents}}]{Hermele2004}%
  \BibitemOpen
  \bibfield  {author} {\bibinfo {author} {\bibfnamefont {M.}~\bibnamefont
  {Hermele}}, \bibinfo {author} {\bibfnamefont {M.~P.}\ \bibnamefont {Fisher}},
  \ and\ \bibinfo {author} {\bibfnamefont {L.}~\bibnamefont {Balents}},\ }\href
  {https://journals.aps.org/prb/abstract/10.1103/PhysRevB.69.064404} {\bibfield
   {journal} {\bibinfo  {journal} {Phys. Rev. B}\ }\textbf {\bibinfo {volume}
  {69}},\ \bibinfo {pages} {064404} (\bibinfo {year} {2004})}\BibitemShut
  {NoStop}%
\bibitem [{\citenamefont {Gingras}\ and\ \citenamefont
  {McClarty}(2014)}]{Gingras2014}%
  \BibitemOpen
  \bibfield  {author} {\bibinfo {author} {\bibfnamefont {M.~J.}\ \bibnamefont
  {Gingras}}\ and\ \bibinfo {author} {\bibfnamefont {P.~A.}\ \bibnamefont
  {McClarty}},\ }\href
  {https://iopscience.iop.org/article/10.1088/0034-4885/77/5/056501} {\bibfield
   {journal} {\bibinfo  {journal} {Rep. Prog. Phys.}\ }\textbf {\bibinfo
  {volume} {77}},\ \bibinfo {pages} {056501} (\bibinfo {year}
  {2014})}\BibitemShut {NoStop}%
\bibitem [{\citenamefont {Pan}\ \emph {et~al.}(2016)\citenamefont {Pan},
  \citenamefont {Laurita}, \citenamefont {Ross}, \citenamefont {Gaulin},\ and\
  \citenamefont {Armitage}}]{pan2016measure}%
  \BibitemOpen
  \bibfield  {author} {\bibinfo {author} {\bibfnamefont {L.}~\bibnamefont
  {Pan}}, \bibinfo {author} {\bibfnamefont {N.}~\bibnamefont {Laurita}},
  \bibinfo {author} {\bibfnamefont {K.~A.}\ \bibnamefont {Ross}}, \bibinfo
  {author} {\bibfnamefont {B.~D.}\ \bibnamefont {Gaulin}}, \ and\ \bibinfo
  {author} {\bibfnamefont {N.}~\bibnamefont {Armitage}},\ }\href
  {https://www.nature.com/articles/nphys3608} {\bibfield  {journal} {\bibinfo
  {journal} {Nat. Phys.}\ }\textbf {\bibinfo {volume} {12}},\ \bibinfo {pages}
  {361} (\bibinfo {year} {2016})}\BibitemShut {NoStop}%
\bibitem [{\citenamefont {Sibille}\ \emph {et~al.}(2018)\citenamefont
  {Sibille}, \citenamefont {Gauthier}, \citenamefont {Yan}, \citenamefont
  {Hatnean}, \citenamefont {Ollivier}, \citenamefont {Winn}, \citenamefont
  {Filges}, \citenamefont {Balakrishnan}, \citenamefont {Kenzelmann},
  \citenamefont {Shannon} \emph {et~al.}}]{sibille2018experimental}%
  \BibitemOpen
  \bibfield  {author} {\bibinfo {author} {\bibfnamefont {R.}~\bibnamefont
  {Sibille}}, \bibinfo {author} {\bibfnamefont {N.}~\bibnamefont {Gauthier}},
  \bibinfo {author} {\bibfnamefont {H.}~\bibnamefont {Yan}}, \bibinfo {author}
  {\bibfnamefont {M.~C.}\ \bibnamefont {Hatnean}}, \bibinfo {author}
  {\bibfnamefont {J.}~\bibnamefont {Ollivier}}, \bibinfo {author}
  {\bibfnamefont {B.}~\bibnamefont {Winn}}, \bibinfo {author} {\bibfnamefont
  {U.}~\bibnamefont {Filges}}, \bibinfo {author} {\bibfnamefont
  {G.}~\bibnamefont {Balakrishnan}}, \bibinfo {author} {\bibfnamefont
  {M.}~\bibnamefont {Kenzelmann}}, \bibinfo {author} {\bibfnamefont
  {N.}~\bibnamefont {Shannon}},  \emph {et~al.},\ }\href
  {https://www.nature.com/articles/s41567-018-0116-x} {\bibfield  {journal}
  {\bibinfo  {journal} {Nat. Phys.}\ }\textbf {\bibinfo {volume} {14}},\
  \bibinfo {pages} {711} (\bibinfo {year} {2018})}\BibitemShut {NoStop}%
\bibitem [{\citenamefont {Gaudet}\ \emph {et~al.}(2019)\citenamefont {Gaudet},
  \citenamefont {Smith}, \citenamefont {Dudemaine}, \citenamefont {Beare},
  \citenamefont {Buhariwalla}, \citenamefont {Butch}, \citenamefont {Stone},
  \citenamefont {Kolesnikov}, \citenamefont {Xu}, \citenamefont {Yahne} \emph
  {et~al.}}]{gaudet2019quantum}%
  \BibitemOpen
  \bibfield  {author} {\bibinfo {author} {\bibfnamefont {J.}~\bibnamefont
  {Gaudet}}, \bibinfo {author} {\bibfnamefont {E.}~\bibnamefont {Smith}},
  \bibinfo {author} {\bibfnamefont {J.}~\bibnamefont {Dudemaine}}, \bibinfo
  {author} {\bibfnamefont {J.}~\bibnamefont {Beare}}, \bibinfo {author}
  {\bibfnamefont {C.}~\bibnamefont {Buhariwalla}}, \bibinfo {author}
  {\bibfnamefont {N.~P.}\ \bibnamefont {Butch}}, \bibinfo {author}
  {\bibfnamefont {M.}~\bibnamefont {Stone}}, \bibinfo {author} {\bibfnamefont
  {A.}~\bibnamefont {Kolesnikov}}, \bibinfo {author} {\bibfnamefont
  {G.}~\bibnamefont {Xu}}, \bibinfo {author} {\bibfnamefont {D.}~\bibnamefont
  {Yahne}},  \emph {et~al.},\ }\href
  {https://journals.aps.org/prl/abstract/10.1103/PhysRevLett.122.187201}
  {\bibfield  {journal} {\bibinfo  {journal} {Phys. Rev. Lett.}\ }\textbf
  {\bibinfo {volume} {122}},\ \bibinfo {pages} {187201} (\bibinfo {year}
  {2019})}\BibitemShut {NoStop}%
\bibitem [{\citenamefont {Gao}\ \emph {et~al.}(2019)\citenamefont {Gao},
  \citenamefont {Chen}, \citenamefont {Tam}, \citenamefont {Huang},
  \citenamefont {Sasmal}, \citenamefont {Adroja}, \citenamefont {Ye},
  \citenamefont {Cao}, \citenamefont {Sala}, \citenamefont {Stone} \emph
  {et~al.}}]{gao2019experimental}%
  \BibitemOpen
  \bibfield  {author} {\bibinfo {author} {\bibfnamefont {B.}~\bibnamefont
  {Gao}}, \bibinfo {author} {\bibfnamefont {T.}~\bibnamefont {Chen}}, \bibinfo
  {author} {\bibfnamefont {D.~W.}\ \bibnamefont {Tam}}, \bibinfo {author}
  {\bibfnamefont {C.-L.}\ \bibnamefont {Huang}}, \bibinfo {author}
  {\bibfnamefont {K.}~\bibnamefont {Sasmal}}, \bibinfo {author} {\bibfnamefont
  {D.~T.}\ \bibnamefont {Adroja}}, \bibinfo {author} {\bibfnamefont
  {F.}~\bibnamefont {Ye}}, \bibinfo {author} {\bibfnamefont {H.}~\bibnamefont
  {Cao}}, \bibinfo {author} {\bibfnamefont {G.}~\bibnamefont {Sala}}, \bibinfo
  {author} {\bibfnamefont {M.~B.}\ \bibnamefont {Stone}},  \emph {et~al.},\
  }\href {https://www.nature.com/articles/s41567-019-0577-6} {\bibfield
  {journal} {\bibinfo  {journal} {Nat. Phys.}\ }\textbf {\bibinfo {volume}
  {15}},\ \bibinfo {pages} {1052} (\bibinfo {year} {2019})}\BibitemShut
  {NoStop}%
\bibitem [{Note1()}]{Note1}%
  \BibitemOpen
  \bibinfo {note} {The terminology for the excitations in QSI differs among
  communities~\cite {Gingras2014}; we adopt the language used by the gauge
  theory literature where the spinon is called an electric charge. Our electric
  charge is referred to as a magnetic monopole in the classical spin ice
  literature and a spinon in the quantum spin liquid literature. Our magnetic
  monopole is also sometimes referred to as a vison in the quantum spin ice
  literature.}\BibitemShut {Stop}%
\bibitem [{\citenamefont {Kogut}(1979)}]{kogut1979introduction}%
  \BibitemOpen
  \bibfield  {author} {\bibinfo {author} {\bibfnamefont {J.~B.}\ \bibnamefont
  {Kogut}},\ }\href
  {https://journals.aps.org/rmp/abstract/10.1103/RevModPhys.51.659} {\bibfield
  {journal} {\bibinfo  {journal} {Rev. Mod. Phys.}\ }\textbf {\bibinfo {volume}
  {51}},\ \bibinfo {pages} {659} (\bibinfo {year} {1979})}\BibitemShut
  {NoStop}%
\bibitem [{\citenamefont {Banerjee}\ \emph {et~al.}(2008)\citenamefont
  {Banerjee}, \citenamefont {Isakov}, \citenamefont {Damle},\ and\
  \citenamefont {Kim}}]{Banerjee2008}%
  \BibitemOpen
  \bibfield  {author} {\bibinfo {author} {\bibfnamefont {A.}~\bibnamefont
  {Banerjee}}, \bibinfo {author} {\bibfnamefont {S.~V.}\ \bibnamefont
  {Isakov}}, \bibinfo {author} {\bibfnamefont {K.}~\bibnamefont {Damle}}, \
  and\ \bibinfo {author} {\bibfnamefont {Y.~B.}\ \bibnamefont {Kim}},\ }\href
  {https://journals.aps.org/prl/abstract/10.1103/PhysRevLett.100.047208}
  {\bibfield  {journal} {\bibinfo  {journal} {Phys. Rev. Lett.}\ }\textbf
  {\bibinfo {volume} {100}},\ \bibinfo {pages} {047208} (\bibinfo {year}
  {2008})}\BibitemShut {NoStop}%
\bibitem [{\citenamefont {Shannon}\ \emph {et~al.}(2012)\citenamefont
  {Shannon}, \citenamefont {Sikora}, \citenamefont {Pollmann}, \citenamefont
  {Penc},\ and\ \citenamefont {Fulde}}]{Shannon2012}%
  \BibitemOpen
  \bibfield  {author} {\bibinfo {author} {\bibfnamefont {N.}~\bibnamefont
  {Shannon}}, \bibinfo {author} {\bibfnamefont {O.}~\bibnamefont {Sikora}},
  \bibinfo {author} {\bibfnamefont {F.}~\bibnamefont {Pollmann}}, \bibinfo
  {author} {\bibfnamefont {K.}~\bibnamefont {Penc}}, \ and\ \bibinfo {author}
  {\bibfnamefont {P.}~\bibnamefont {Fulde}},\ }\href
  {https://journals.aps.org/prl/abstract/10.1103/PhysRevLett.108.067204}
  {\bibfield  {journal} {\bibinfo  {journal} {Phys. Rev. Lett.}\ }\textbf
  {\bibinfo {volume} {108}},\ \bibinfo {pages} {067204} (\bibinfo {year}
  {2012})}\BibitemShut {NoStop}%
\bibitem [{\citenamefont {Benton}\ \emph {et~al.}(2012)\citenamefont {Benton},
  \citenamefont {Sikora},\ and\ \citenamefont {Shannon}}]{Benton2012}%
  \BibitemOpen
  \bibfield  {author} {\bibinfo {author} {\bibfnamefont {O.}~\bibnamefont
  {Benton}}, \bibinfo {author} {\bibfnamefont {O.}~\bibnamefont {Sikora}}, \
  and\ \bibinfo {author} {\bibfnamefont {N.}~\bibnamefont {Shannon}},\ }\href
  {https://journals.aps.org/prb/abstract/10.1103/PhysRevB.86.075154} {\bibfield
   {journal} {\bibinfo  {journal} {Phys. Rev. B}\ }\textbf {\bibinfo {volume}
  {86}},\ \bibinfo {pages} {075154} (\bibinfo {year} {2012})}\BibitemShut
  {NoStop}%
\bibitem [{\citenamefont {Kato}\ and\ \citenamefont {Onoda}(2015)}]{Kato2015}%
  \BibitemOpen
  \bibfield  {author} {\bibinfo {author} {\bibfnamefont {Y.}~\bibnamefont
  {Kato}}\ and\ \bibinfo {author} {\bibfnamefont {S.}~\bibnamefont {Onoda}},\
  }\href {https://journals.aps.org/prl/abstract/10.1103/PhysRevLett.115.077202}
  {\bibfield  {journal} {\bibinfo  {journal} {Phys. Rev. Lett.}\ }\textbf
  {\bibinfo {volume} {115}},\ \bibinfo {pages} {077202} (\bibinfo {year}
  {2015})}\BibitemShut {NoStop}%
\bibitem [{\citenamefont {Szab\'o}\ and\ \citenamefont
  {Castelnovo}(2019)}]{Szabo2019a}%
  \BibitemOpen
  \bibfield  {author} {\bibinfo {author} {\bibfnamefont {A.}~\bibnamefont
  {Szab\'o}}\ and\ \bibinfo {author} {\bibfnamefont {C.}~\bibnamefont
  {Castelnovo}},\ }\href {\doibase 10.1103/PhysRevB.100.014417} {\bibfield
  {journal} {\bibinfo  {journal} {Phys. Rev. B}\ }\textbf {\bibinfo {volume}
  {100}},\ \bibinfo {pages} {014417} (\bibinfo {year} {2019})}\BibitemShut
  {NoStop}%
\bibitem [{\citenamefont {Kwasigroch}(2020)}]{PhysRevB.102.125113}%
  \BibitemOpen
  \bibfield  {author} {\bibinfo {author} {\bibfnamefont {M.~P.}\ \bibnamefont
  {Kwasigroch}},\ }\href {\doibase 10.1103/PhysRevB.102.125113} {\bibfield
  {journal} {\bibinfo  {journal} {Phys. Rev. B}\ }\textbf {\bibinfo {volume}
  {102}},\ \bibinfo {pages} {125113} (\bibinfo {year} {2020})}\BibitemShut
  {NoStop}%
\bibitem [{\citenamefont {Bramwell}\ and\ \citenamefont
  {Gingras}(2001)}]{Bramwell2001}%
  \BibitemOpen
  \bibfield  {author} {\bibinfo {author} {\bibfnamefont {S.~T.}\ \bibnamefont
  {Bramwell}}\ and\ \bibinfo {author} {\bibfnamefont {M.~J.}\ \bibnamefont
  {Gingras}},\ }\href {https://science.sciencemag.org/content/294/5546/1495}
  {\bibfield  {journal} {\bibinfo  {journal} {Science}\ }\textbf {\bibinfo
  {volume} {294}},\ \bibinfo {pages} {1495} (\bibinfo {year}
  {2001})}\BibitemShut {NoStop}%
\bibitem [{\citenamefont {Anderson}(1956)}]{Anderson1956}%
  \BibitemOpen
  \bibfield  {author} {\bibinfo {author} {\bibfnamefont {P.~W.}\ \bibnamefont
  {Anderson}},\ }\href {\doibase 10.1103/PhysRev.102.1008} {\bibfield
  {journal} {\bibinfo  {journal} {Phys. Rev.}\ }\textbf {\bibinfo {volume}
  {102}},\ \bibinfo {pages} {1008} (\bibinfo {year} {1956})}\BibitemShut
  {NoStop}%
\bibitem [{\citenamefont {Castelnovo}\ \emph {et~al.}(2012)\citenamefont
  {Castelnovo}, \citenamefont {Moessner},\ and\ \citenamefont
  {Sondhi}}]{Castelnovo2012}%
  \BibitemOpen
  \bibfield  {author} {\bibinfo {author} {\bibfnamefont {C.}~\bibnamefont
  {Castelnovo}}, \bibinfo {author} {\bibfnamefont {R.}~\bibnamefont
  {Moessner}}, \ and\ \bibinfo {author} {\bibfnamefont {S.~L.}\ \bibnamefont
  {Sondhi}},\ }\href
  {https://www.annualreviews.org/doi/10.1146/annurev-conmatphys-020911-125058}
  {\bibfield  {journal} {\bibinfo  {journal} {Annu. Rev. Condens. Matter
  Phys.}\ }\textbf {\bibinfo {volume} {3}},\ \bibinfo {pages} {35} (\bibinfo
  {year} {2012})}\BibitemShut {NoStop}%
\bibitem [{\citenamefont {Huang}\ \emph {et~al.}(2018)\citenamefont {Huang},
  \citenamefont {Deng}, \citenamefont {Wan},\ and\ \citenamefont
  {Meng}}]{Huang2018}%
  \BibitemOpen
  \bibfield  {author} {\bibinfo {author} {\bibfnamefont {C.-J.}\ \bibnamefont
  {Huang}}, \bibinfo {author} {\bibfnamefont {Y.}~\bibnamefont {Deng}},
  \bibinfo {author} {\bibfnamefont {Y.}~\bibnamefont {Wan}}, \ and\ \bibinfo
  {author} {\bibfnamefont {Z.~Y.}\ \bibnamefont {Meng}},\ }\href
  {https://journals.aps.org/prl/abstract/10.1103/PhysRevLett.120.167202}
  {\bibfield  {journal} {\bibinfo  {journal} {Phys. Rev. Lett.}\ }\textbf
  {\bibinfo {volume} {120}},\ \bibinfo {pages} {167202} (\bibinfo {year}
  {2018})}\BibitemShut {NoStop}%
\bibitem [{\citenamefont {Ross}\ \emph {et~al.}(2011)\citenamefont {Ross},
  \citenamefont {Savary}, \citenamefont {Gaulin},\ and\ \citenamefont
  {Balents}}]{Ross2011}%
  \BibitemOpen
  \bibfield  {author} {\bibinfo {author} {\bibfnamefont {K.~A.}\ \bibnamefont
  {Ross}}, \bibinfo {author} {\bibfnamefont {L.}~\bibnamefont {Savary}},
  \bibinfo {author} {\bibfnamefont {B.~D.}\ \bibnamefont {Gaulin}}, \ and\
  \bibinfo {author} {\bibfnamefont {L.}~\bibnamefont {Balents}},\ }\href
  {https://journals.aps.org/prx/abstract/10.1103/PhysRevX.1.021002} {\bibfield
  {journal} {\bibinfo  {journal} {Phys. Rev. X}\ }\textbf {\bibinfo {volume}
  {1}},\ \bibinfo {pages} {021002} (\bibinfo {year} {2011})}\BibitemShut
  {NoStop}%
\bibitem [{\citenamefont {Savary}\ and\ \citenamefont
  {Balents}(2012)}]{Savary2012}%
  \BibitemOpen
  \bibfield  {author} {\bibinfo {author} {\bibfnamefont {L.}~\bibnamefont
  {Savary}}\ and\ \bibinfo {author} {\bibfnamefont {L.}~\bibnamefont
  {Balents}},\ }\href
  {https://journals.aps.org/prl/abstract/10.1103/PhysRevLett.108.037202}
  {\bibfield  {journal} {\bibinfo  {journal} {Phys. Rev. Lett.}\ }\textbf
  {\bibinfo {volume} {108}},\ \bibinfo {pages} {037202} (\bibinfo {year}
  {2012})}\BibitemShut {NoStop}%
\bibitem [{\citenamefont {Lee}\ \emph {et~al.}(2012)\citenamefont {Lee},
  \citenamefont {Onoda},\ and\ \citenamefont {Balents}}]{Lee2012}%
  \BibitemOpen
  \bibfield  {author} {\bibinfo {author} {\bibfnamefont {S.}~\bibnamefont
  {Lee}}, \bibinfo {author} {\bibfnamefont {S.}~\bibnamefont {Onoda}}, \ and\
  \bibinfo {author} {\bibfnamefont {L.}~\bibnamefont {Balents}},\ }\href
  {\doibase 10.1103/PhysRevB.86.104412} {\bibfield  {journal} {\bibinfo
  {journal} {Phys. Rev. B}\ }\textbf {\bibinfo {volume} {86}},\ \bibinfo
  {pages} {104412} (\bibinfo {year} {2012})}\BibitemShut {NoStop}%
\bibitem [{\citenamefont {Rau}\ and\ \citenamefont {Gingras}(2019)}]{Rau2019}%
  \BibitemOpen
  \bibfield  {author} {\bibinfo {author} {\bibfnamefont {J.~G.}\ \bibnamefont
  {Rau}}\ and\ \bibinfo {author} {\bibfnamefont {M.~J.}\ \bibnamefont
  {Gingras}},\ }\href
  {https://www.annualreviews.org/doi/abs/10.1146/annurev-conmatphys-022317-110520}
  {\bibfield  {journal} {\bibinfo  {journal} {Annu. Rev. Condens. Matter
  Phys.}\ }\textbf {\bibinfo {volume} {10}},\ \bibinfo {pages} {357} (\bibinfo
  {year} {2019})}\BibitemShut {NoStop}%
\bibitem [{\citenamefont {Moessner}\ and\ \citenamefont
  {Raman}(2011)}]{Moessner2011}%
  \BibitemOpen
  \bibfield  {author} {\bibinfo {author} {\bibfnamefont {R.}~\bibnamefont
  {Moessner}}\ and\ \bibinfo {author} {\bibfnamefont {K.~S.}\ \bibnamefont
  {Raman}},\ }in\ \href@noop {} {\emph {\bibinfo {booktitle} {Introduction to
  Frustrated Magnetism}}}\ (\bibinfo  {publisher} {Springer},\ \bibinfo {year}
  {2011})\ pp.\ \bibinfo {pages} {437--479}\BibitemShut {NoStop}%
\bibitem [{\citenamefont {King}\ \emph {et~al.}(2020)\citenamefont {King},
  \citenamefont {Nisoli}, \citenamefont {Dahl}, \citenamefont
  {Poulin-Lamarre},\ and\ \citenamefont {Lopez-Bezanilla}}]{King2020}%
  \BibitemOpen
  \bibfield  {author} {\bibinfo {author} {\bibfnamefont {A.~D.}\ \bibnamefont
  {King}}, \bibinfo {author} {\bibfnamefont {C.}~\bibnamefont {Nisoli}},
  \bibinfo {author} {\bibfnamefont {E.~D.}\ \bibnamefont {Dahl}}, \bibinfo
  {author} {\bibfnamefont {G.}~\bibnamefont {Poulin-Lamarre}}, \ and\ \bibinfo
  {author} {\bibfnamefont {A.}~\bibnamefont {Lopez-Bezanilla}},\ }\href
  {https://arxiv.org/abs/2007.10555} {\bibfield  {journal} {\bibinfo  {journal}
  {arXiv preprint arXiv:2007.10555}\ } (\bibinfo {year} {2020})}\BibitemShut
  {NoStop}%
\bibitem [{\citenamefont {Bernien}\ \emph {et~al.}(2017)\citenamefont
  {Bernien}, \citenamefont {Schwartz}, \citenamefont {Keesling}, \citenamefont
  {Levine}, \citenamefont {Omran}, \citenamefont {Pichler}, \citenamefont
  {Choi}, \citenamefont {Zibrov}, \citenamefont {Endres}, \citenamefont
  {Greiner}, \citenamefont {Vuleti{\'{c}}},\ and\ \citenamefont
  {Lukin}}]{Bernien:2017to}%
  \BibitemOpen
  \bibfield  {author} {\bibinfo {author} {\bibfnamefont {H.}~\bibnamefont
  {Bernien}}, \bibinfo {author} {\bibfnamefont {S.}~\bibnamefont {Schwartz}},
  \bibinfo {author} {\bibfnamefont {A.}~\bibnamefont {Keesling}}, \bibinfo
  {author} {\bibfnamefont {H.}~\bibnamefont {Levine}}, \bibinfo {author}
  {\bibfnamefont {A.}~\bibnamefont {Omran}}, \bibinfo {author} {\bibfnamefont
  {H.}~\bibnamefont {Pichler}}, \bibinfo {author} {\bibfnamefont
  {S.}~\bibnamefont {Choi}}, \bibinfo {author} {\bibfnamefont {A.~S.}\
  \bibnamefont {Zibrov}}, \bibinfo {author} {\bibfnamefont {M.}~\bibnamefont
  {Endres}}, \bibinfo {author} {\bibfnamefont {M.}~\bibnamefont {Greiner}},
  \bibinfo {author} {\bibfnamefont {V.}~\bibnamefont {Vuleti{\'{c}}}}, \ and\
  \bibinfo {author} {\bibfnamefont {M.~D.}\ \bibnamefont {Lukin}},\ }\href
  {\doibase 10.1038/nature24622} {\bibfield  {journal} {\bibinfo  {journal}
  {Nature}\ }\textbf {\bibinfo {volume} {551}},\ \bibinfo {pages} {579}
  (\bibinfo {year} {2017})}\BibitemShut {NoStop}%
\bibitem [{\citenamefont {Pagano}\ \emph {et~al.}(2020)\citenamefont {Pagano},
  \citenamefont {Bapat}, \citenamefont {Becker}, \citenamefont {Collins},
  \citenamefont {De}, \citenamefont {Hess}, \citenamefont {Kaplan},
  \citenamefont {Kyprianidis}, \citenamefont {Tan}, \citenamefont {Baldwin},
  \citenamefont {Brady}, \citenamefont {Deshpande}, \citenamefont {Liu},
  \citenamefont {Jordan}, \citenamefont {Gorshkov},\ and\ \citenamefont
  {Monroe}}]{Pagano:2020wp}%
  \BibitemOpen
  \bibfield  {author} {\bibinfo {author} {\bibfnamefont {G.}~\bibnamefont
  {Pagano}}, \bibinfo {author} {\bibfnamefont {A.}~\bibnamefont {Bapat}},
  \bibinfo {author} {\bibfnamefont {P.}~\bibnamefont {Becker}}, \bibinfo
  {author} {\bibfnamefont {K.~S.}\ \bibnamefont {Collins}}, \bibinfo {author}
  {\bibfnamefont {A.}~\bibnamefont {De}}, \bibinfo {author} {\bibfnamefont
  {P.~W.}\ \bibnamefont {Hess}}, \bibinfo {author} {\bibfnamefont {H.~B.}\
  \bibnamefont {Kaplan}}, \bibinfo {author} {\bibfnamefont {A.}~\bibnamefont
  {Kyprianidis}}, \bibinfo {author} {\bibfnamefont {W.~L.}\ \bibnamefont
  {Tan}}, \bibinfo {author} {\bibfnamefont {C.}~\bibnamefont {Baldwin}},
  \bibinfo {author} {\bibfnamefont {L.~T.}\ \bibnamefont {Brady}}, \bibinfo
  {author} {\bibfnamefont {A.}~\bibnamefont {Deshpande}}, \bibinfo {author}
  {\bibfnamefont {F.}~\bibnamefont {Liu}}, \bibinfo {author} {\bibfnamefont
  {S.}~\bibnamefont {Jordan}}, \bibinfo {author} {\bibfnamefont {A.~V.}\
  \bibnamefont {Gorshkov}}, \ and\ \bibinfo {author} {\bibfnamefont
  {C.}~\bibnamefont {Monroe}},\ }\href {\doibase 10.1073/pnas.2006373117}
  {\bibfield  {journal} {\bibinfo  {journal} {Proc. Natl. Acad. Sci. U.S.A.}\
  }\textbf {\bibinfo {volume} {117}},\ \bibinfo {pages} {25396} (\bibinfo
  {year} {2020})}\BibitemShut {NoStop}%
\bibitem [{\citenamefont {Rokhsar}\ and\ \citenamefont
  {Kivelson}(1988)}]{Rokhsar1988}%
  \BibitemOpen
  \bibfield  {author} {\bibinfo {author} {\bibfnamefont {D.~S.}\ \bibnamefont
  {Rokhsar}}\ and\ \bibinfo {author} {\bibfnamefont {S.~A.}\ \bibnamefont
  {Kivelson}},\ }\href
  {https://journals.aps.org/prl/abstract/10.1103/PhysRevLett.61.2376}
  {\bibfield  {journal} {\bibinfo  {journal} {Phys. Rev. Lett.}\ }\textbf
  {\bibinfo {volume} {61}},\ \bibinfo {pages} {2376} (\bibinfo {year}
  {1988})}\BibitemShut {NoStop}%
\bibitem [{\citenamefont {Kwasigroch}\ \emph {et~al.}(2017)\citenamefont
  {Kwasigroch}, \citenamefont {Dou{\c{c}}ot},\ and\ \citenamefont
  {Castelnovo}}]{Kwasigroch2017}%
  \BibitemOpen
  \bibfield  {author} {\bibinfo {author} {\bibfnamefont {M.}~\bibnamefont
  {Kwasigroch}}, \bibinfo {author} {\bibfnamefont {B.}~\bibnamefont
  {Dou{\c{c}}ot}}, \ and\ \bibinfo {author} {\bibfnamefont {C.}~\bibnamefont
  {Castelnovo}},\ }\href
  {https://journals.aps.org/prb/abstract/10.1103/PhysRevB.95.134439} {\bibfield
   {journal} {\bibinfo  {journal} {Phys. Rev. B}\ }\textbf {\bibinfo {volume}
  {95}},\ \bibinfo {pages} {134439} (\bibinfo {year} {2017})}\BibitemShut
  {NoStop}%
\bibitem [{\citenamefont {Jers{\'a}k}\ \emph {et~al.}(1983)\citenamefont
  {Jers{\'a}k}, \citenamefont {Neuhaus},\ and\ \citenamefont
  {Zerwas}}]{Jersak1983}%
  \BibitemOpen
  \bibfield  {author} {\bibinfo {author} {\bibfnamefont {J.}~\bibnamefont
  {Jers{\'a}k}}, \bibinfo {author} {\bibfnamefont {T.}~\bibnamefont {Neuhaus}},
  \ and\ \bibinfo {author} {\bibfnamefont {P.~M.}\ \bibnamefont {Zerwas}},\
  }\href {https://www.sciencedirect.com/science/article/pii/0370269383901156}
  {\bibfield  {journal} {\bibinfo  {journal} {Phys. Lett. B}\ }\textbf
  {\bibinfo {volume} {133}},\ \bibinfo {pages} {103} (\bibinfo {year}
  {1983})}\BibitemShut {NoStop}%
\bibitem [{\citenamefont {Gardner}\ \emph {et~al.}(2010)\citenamefont
  {Gardner}, \citenamefont {Gingras},\ and\ \citenamefont
  {Greedan}}]{Gardnner2010}%
  \BibitemOpen
  \bibfield  {author} {\bibinfo {author} {\bibfnamefont {J.~S.}\ \bibnamefont
  {Gardner}}, \bibinfo {author} {\bibfnamefont {M.~J.~P.}\ \bibnamefont
  {Gingras}}, \ and\ \bibinfo {author} {\bibfnamefont {J.~E.}\ \bibnamefont
  {Greedan}},\ }\href {\doibase 10.1103/RevModPhys.82.53} {\bibfield  {journal}
  {\bibinfo  {journal} {Rev. Mod. Phys.}\ }\textbf {\bibinfo {volume} {82}},\
  \bibinfo {pages} {53} (\bibinfo {year} {2010})}\BibitemShut {NoStop}%
\bibitem [{\citenamefont {Morampudi}\ \emph {et~al.}(2020)\citenamefont
  {Morampudi}, \citenamefont {Wilczek},\ and\ \citenamefont
  {Laumann}}]{Morampudi2020}%
  \BibitemOpen
  \bibfield  {author} {\bibinfo {author} {\bibfnamefont {S.~C.}\ \bibnamefont
  {Morampudi}}, \bibinfo {author} {\bibfnamefont {F.}~\bibnamefont {Wilczek}},
  \ and\ \bibinfo {author} {\bibfnamefont {C.~R.}\ \bibnamefont {Laumann}},\
  }\href {https://journals.aps.org/prl/abstract/10.1103/PhysRevLett.124.097204}
  {\bibfield  {journal} {\bibinfo  {journal} {Phys. Rev. Lett.}\ }\textbf
  {\bibinfo {volume} {124}},\ \bibinfo {pages} {097204} (\bibinfo {year}
  {2020})}\BibitemShut {NoStop}%
\bibitem [{\citenamefont {Preskill}(2018)}]{Preskill2018}%
  \BibitemOpen
  \bibfield  {author} {\bibinfo {author} {\bibfnamefont {J.}~\bibnamefont
  {Preskill}},\ }\href {\doibase 10.22331/q-2018-08-06-79} {\bibfield
  {journal} {\bibinfo  {journal} {{Quantum}}\ }\textbf {\bibinfo {volume}
  {2}},\ \bibinfo {pages} {79} (\bibinfo {year} {2018})}\BibitemShut {NoStop}%
\bibitem [{\citenamefont {Glaetzle}\ \emph {et~al.}(2014)\citenamefont
  {Glaetzle}, \citenamefont {Dalmonte}, \citenamefont {Nath}, \citenamefont
  {Rousochatzakis}, \citenamefont {Moessner},\ and\ \citenamefont
  {Zoller}}]{Glaetzle2014}%
  \BibitemOpen
  \bibfield  {author} {\bibinfo {author} {\bibfnamefont {A.~W.}\ \bibnamefont
  {Glaetzle}}, \bibinfo {author} {\bibfnamefont {M.}~\bibnamefont {Dalmonte}},
  \bibinfo {author} {\bibfnamefont {R.}~\bibnamefont {Nath}}, \bibinfo {author}
  {\bibfnamefont {I.}~\bibnamefont {Rousochatzakis}}, \bibinfo {author}
  {\bibfnamefont {R.}~\bibnamefont {Moessner}}, \ and\ \bibinfo {author}
  {\bibfnamefont {P.}~\bibnamefont {Zoller}},\ }\href
  {https://journals.aps.org/prx/abstract/10.1103/PhysRevX.4.041037} {\bibfield
  {journal} {\bibinfo  {journal} {Phys. Rev. X}\ }\textbf {\bibinfo {volume}
  {4}},\ \bibinfo {pages} {041037} (\bibinfo {year} {2014})}\BibitemShut
  {NoStop}%
\bibitem [{\citenamefont {Martinez}\ \emph {et~al.}(2016)\citenamefont
  {Martinez}, \citenamefont {Muschik}, \citenamefont {Schindler}, \citenamefont
  {Nigg}, \citenamefont {Erhard}, \citenamefont {Heyl}, \citenamefont {Hauke},
  \citenamefont {Dalmonte}, \citenamefont {Monz}, \citenamefont {Zoller} \emph
  {et~al.}}]{Martinez2016}%
  \BibitemOpen
  \bibfield  {author} {\bibinfo {author} {\bibfnamefont {E.~A.}\ \bibnamefont
  {Martinez}}, \bibinfo {author} {\bibfnamefont {C.~A.}\ \bibnamefont
  {Muschik}}, \bibinfo {author} {\bibfnamefont {P.}~\bibnamefont {Schindler}},
  \bibinfo {author} {\bibfnamefont {D.}~\bibnamefont {Nigg}}, \bibinfo {author}
  {\bibfnamefont {A.}~\bibnamefont {Erhard}}, \bibinfo {author} {\bibfnamefont
  {M.}~\bibnamefont {Heyl}}, \bibinfo {author} {\bibfnamefont {P.}~\bibnamefont
  {Hauke}}, \bibinfo {author} {\bibfnamefont {M.}~\bibnamefont {Dalmonte}},
  \bibinfo {author} {\bibfnamefont {T.}~\bibnamefont {Monz}}, \bibinfo {author}
  {\bibfnamefont {P.}~\bibnamefont {Zoller}},  \emph {et~al.},\ }\href
  {https://www.nature.com/articles/nature18318} {\bibfield  {journal} {\bibinfo
   {journal} {Nature}\ }\textbf {\bibinfo {volume} {534}},\ \bibinfo {pages}
  {516} (\bibinfo {year} {2016})}\BibitemShut {NoStop}%
\bibitem [{\citenamefont {Yang}\ \emph {et~al.}(2020)\citenamefont {Yang},
  \citenamefont {Sun}, \citenamefont {Ott}, \citenamefont {Wang}, \citenamefont
  {Zache}, \citenamefont {Halimeh}, \citenamefont {Yuan}, \citenamefont
  {Hauke},\ and\ \citenamefont {Pan}}]{Yang2020}%
  \BibitemOpen
  \bibfield  {author} {\bibinfo {author} {\bibfnamefont {B.}~\bibnamefont
  {Yang}}, \bibinfo {author} {\bibfnamefont {H.}~\bibnamefont {Sun}}, \bibinfo
  {author} {\bibfnamefont {R.}~\bibnamefont {Ott}}, \bibinfo {author}
  {\bibfnamefont {H.-Y.}\ \bibnamefont {Wang}}, \bibinfo {author}
  {\bibfnamefont {T.~V.}\ \bibnamefont {Zache}}, \bibinfo {author}
  {\bibfnamefont {J.~C.}\ \bibnamefont {Halimeh}}, \bibinfo {author}
  {\bibfnamefont {Z.-S.}\ \bibnamefont {Yuan}}, \bibinfo {author}
  {\bibfnamefont {P.}~\bibnamefont {Hauke}}, \ and\ \bibinfo {author}
  {\bibfnamefont {J.-W.}\ \bibnamefont {Pan}},\ }\href
  {https://www.nature.com/articles/s41586-020-2910-8} {\bibfield  {journal}
  {\bibinfo  {journal} {Nature}\ }\textbf {\bibinfo {volume} {587}},\ \bibinfo
  {pages} {392} (\bibinfo {year} {2020})}\BibitemShut {NoStop}%
\bibitem [{\citenamefont {Wilczynska}\ \emph {et~al.}(2020)\citenamefont
  {Wilczynska}, \citenamefont {Webb}, \citenamefont {Bainbridge}, \citenamefont
  {Barrow}, \citenamefont {Bosman}, \citenamefont {Carswell}, \citenamefont
  {D{\k{a}}browski}, \citenamefont {Dumont}, \citenamefont {Lee}, \citenamefont
  {Leite} \emph {et~al.}}]{Wilczynska2020}%
  \BibitemOpen
  \bibfield  {author} {\bibinfo {author} {\bibfnamefont {M.~R.}\ \bibnamefont
  {Wilczynska}}, \bibinfo {author} {\bibfnamefont {J.~K.}\ \bibnamefont
  {Webb}}, \bibinfo {author} {\bibfnamefont {M.}~\bibnamefont {Bainbridge}},
  \bibinfo {author} {\bibfnamefont {J.~D.}\ \bibnamefont {Barrow}}, \bibinfo
  {author} {\bibfnamefont {S.~E.}\ \bibnamefont {Bosman}}, \bibinfo {author}
  {\bibfnamefont {R.~F.}\ \bibnamefont {Carswell}}, \bibinfo {author}
  {\bibfnamefont {M.~P.}\ \bibnamefont {D{\k{a}}browski}}, \bibinfo {author}
  {\bibfnamefont {V.}~\bibnamefont {Dumont}}, \bibinfo {author} {\bibfnamefont
  {C.-C.}\ \bibnamefont {Lee}}, \bibinfo {author} {\bibfnamefont {A.~C.}\
  \bibnamefont {Leite}},  \emph {et~al.},\ }\href
  {https://advances.sciencemag.org/content/6/17/eaay9672} {\bibfield  {journal}
  {\bibinfo  {journal} {Sci. Adv.}\ }\textbf {\bibinfo {volume} {6}},\ \bibinfo
  {pages} {eaay9672} (\bibinfo {year} {2020})}\BibitemShut {NoStop}%
\bibitem [{\citenamefont {Sommerfeld}(1916)}]{Sommerfeld1916}%
  \BibitemOpen
  \bibfield  {author} {\bibinfo {author} {\bibfnamefont {A.}~\bibnamefont
  {Sommerfeld}},\ }\href
  {https://onlinelibrary.wiley.com/doi/abs/10.1002/andp.19163561702} {\bibfield
   {journal} {\bibinfo  {journal} {Ann. Phys.}\ }\textbf {\bibinfo {volume}
  {356}},\ \bibinfo {pages} {1} (\bibinfo {year} {1916})}\BibitemShut {NoStop}%
\bibitem [{\citenamefont {Sikora}\ \emph {et~al.}(2011)\citenamefont {Sikora},
  \citenamefont {Shannon}, \citenamefont {Pollmann}, \citenamefont {Penc},\
  and\ \citenamefont {Fulde}}]{PhysRevB.84.115129}%
  \BibitemOpen
  \bibfield  {author} {\bibinfo {author} {\bibfnamefont {O.}~\bibnamefont
  {Sikora}}, \bibinfo {author} {\bibfnamefont {N.}~\bibnamefont {Shannon}},
  \bibinfo {author} {\bibfnamefont {F.}~\bibnamefont {Pollmann}}, \bibinfo
  {author} {\bibfnamefont {K.}~\bibnamefont {Penc}}, \ and\ \bibinfo {author}
  {\bibfnamefont {P.}~\bibnamefont {Fulde}},\ }\href {\doibase
  10.1103/PhysRevB.84.115129} {\bibfield  {journal} {\bibinfo  {journal} {Phys.
  Rev. B}\ }\textbf {\bibinfo {volume} {84}},\ \bibinfo {pages} {115129}
  (\bibinfo {year} {2011})}\BibitemShut {NoStop}%
\bibitem [{\citenamefont {Sandvik}(2010)}]{Sandvik2010}%
  \BibitemOpen
  \bibfield  {author} {\bibinfo {author} {\bibfnamefont {A.~W.}\ \bibnamefont
  {Sandvik}},\ }in\ \href {https://aip.scitation.org/doi/abs/10.1063/1.3518900}
  {\emph {\bibinfo {booktitle} {AIP Conference Proceedings}}},\ Vol.\ \bibinfo
  {volume} {1297}\ (\bibinfo {organization} {American Institute of Physics},\
  \bibinfo {year} {2010})\ pp.\ \bibinfo {pages} {135--338}\BibitemShut
  {NoStop}%
\bibitem [{\citenamefont {Chen}(2016)}]{Chen2016}%
  \BibitemOpen
  \bibfield  {author} {\bibinfo {author} {\bibfnamefont {G.}~\bibnamefont
  {Chen}},\ }\href {\doibase 10.1103/PhysRevB.94.205107} {\bibfield  {journal}
  {\bibinfo  {journal} {Phys. Rev. B}\ }\textbf {\bibinfo {volume} {94}},\
  \bibinfo {pages} {205107} (\bibinfo {year} {2016})}\BibitemShut {NoStop}%
\end{thebibliography}%

\pagebreak
\onecolumngrid
\pagebreak

\section*{Supplemental Material}

\textbf{Exact Diagonalization Methods} Our results are derived by analyzing the low-energy spectra measured from large-scale exact diagonalization (ED) of the microscopic Hamiltonians given by Eqn.~\eqref{eqn:QSIeff} at different values of $\zeta$ and $\mu$ in Eqn.~\eqref{eqn:QSIp}. 
Crucially, we work with periodic boundary conditions and project into the constrained Hilbert space strictly satisfying the ice rules (Gauss law). 
This provides access to much larger systems and enables us to exploit the strict conservation of the electric flux through the boundaries to measure the elementary charge.
The Pauling estimate for the entropy per spin of the ice manifold is approximately $\frac{1}{2} \ln \frac{3}{2} \approx 0.2$, which is much less than that of the unconstrained spin entropy, $\ln 2 \approx 0.7$; 
this underlies our ability to reach large systems with up to $N=96$ spins. 

\begin{figure}[b!]
\centering
    \includegraphics[width=.35\textwidth]{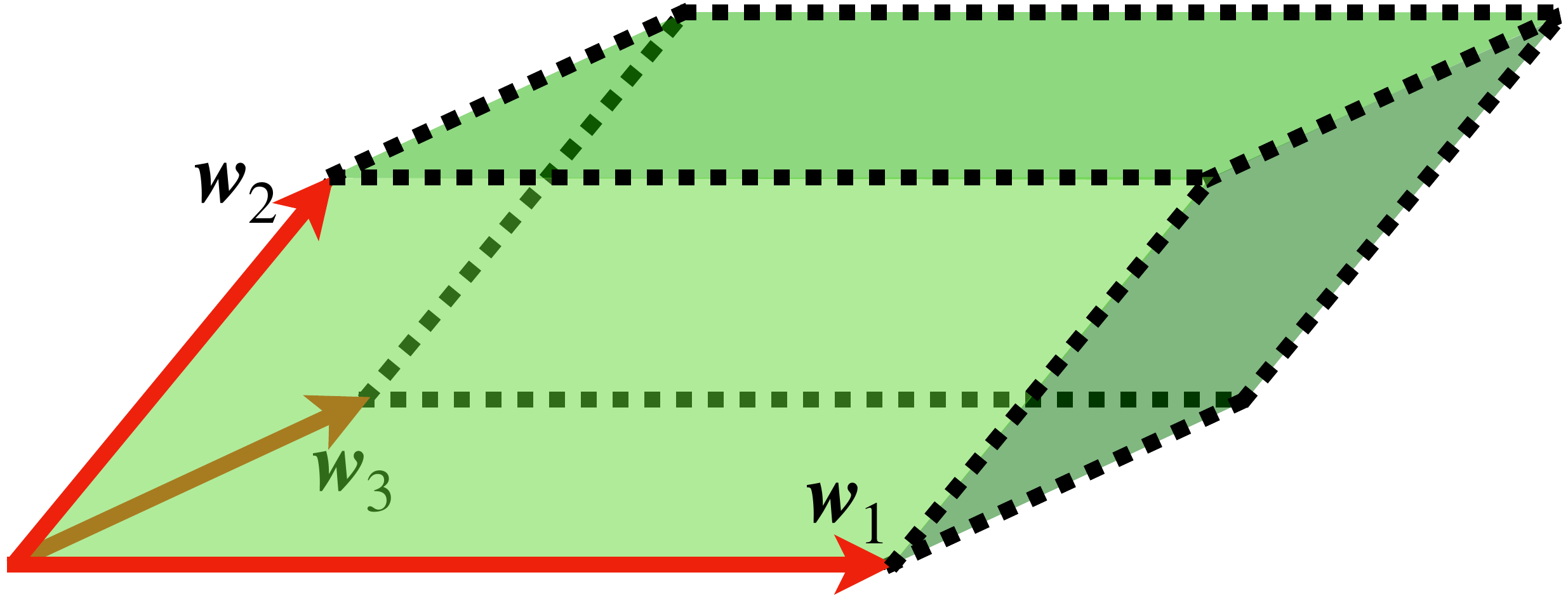}
    \caption{A parallelepiped which is the canonical volume the periodic chunk of pyrochlore lattice is embedded in. The wrapping vectors $\bm{w_{1}}$, $\bm{w_{2}}$, and $\bm{w_{3}}$ characterizes the shape of the lattice.}
    \label{fig:parallelepiped}
\end{figure}

The ED data is collected using various shapes of the pyrochlore lattice with 56 to 96 spins. 
The shape of the lattice is determined by three wrapping vectors, $\bm{w_{1}}$, $\bm{w_{2}}$, and $\bm{w_{3}}$ which define the canonical volume the lattice is embedded in, as shown in Fig.~\ref{fig:parallelepiped}.
The ED data shown in Figs. \ref{fig:eQEDandAlpha}c and \ref{fig:e2_and_c_data} in the main text comes from an extensive range of wrapping vectors corresponding to 180 unique periodic units.
These different shapes are sampled by generating a random $3\times 3$ integer matrix whose column vectors are the three wrapping vectors. 
From these three wrapping vectors, we check the number of unit cells they correspond to and ensure each hexagonal plaquette is made up of 6 unique edges of the lattice (small shapes can have a hexagonal plaquette made up of the same edge multiple times due to periodic boundary conditions).
In (3+1)D, even 96 spins corresponds to a shape with fairly small linear dimension and thus there is a noticeable shape dependence of the measurements. 
Our central estimates for $e_\mathrm{QSI}$ and $c_{\mathrm{QSI}}$ are extracted by best fit to the relevant spectral data across all of the collected shapes. 
The error bars represent the standard deviation of $e_{\mathrm{QSI}}$ and $c_{\mathrm{QSI}}$ among its shape-dependent variations at a fixed ($\mu$,$\zeta$).

Given a particular finite-size periodic geometry, specified by $\bm{w}_i$, with $N$ spins, it is computationally prohibitive to generate all $2^N$ spin states and filter them down to those which satisfy the ice rules.
Rather, we generate the constrained Hilbert spaces in two steps. 
First, from a uniform reference state satisfying the ice rule, we introduce a charge-anticharge pair and cause the charge to randomly walk until it reannihilates with the anticharge. 
This 'worm' algorithm find states lying in distinct electric flux sectors $\bm{\phi} \in \mathbb{Z}^3$ whenever the path it takes winds through a periodic boundary.
By repeating this many times, we find individual states in many different flux sectors.
Second, given a state in a particular flux sector generated by running the worm, we generate all states within the sector by exhaustive traversal of the state space generated by the local ring exchange moves $W_{\hexagon}$.

We found a small number of shapes for which the local ring exchange is non-ergodic  within the flux sectors. 
We identify these by running the worm many times and checking whether the states it finds with the same electric flux are in fact connected under $W_{\hexagon}$. 
The non-ergodic shapes we found are all made from simple cubic unit cells of the diamond lattice stacked in a line. 
In fact, they are the same shapes considered in Ref.~\cite{PhysRevB.84.115129}, which studied the quantum dimer model on the diamond lattice and offered an incomplete understanding of the quantum numbers corresponding to the different disconnected sectors. 
We excluded these quasi-one dimensional shapes from our results. 

Finally, the periodic boundary conditions allow us to decompose each of the flux sectors $\mathcal{H}_{\bm{\phi}}$ into momentum $\bm{k} = (k_1, k_2, k_3)$ subspaces $\mathcal{H}_{\bm{\phi}, \bm{k}}$. 
This projection is accomplished using standard symmetry projection techniques adapted to the constrained Hilbert space \cite{Sandvik2010}. 
Ultimately, our estimates of $c_{\mathrm{QSI}}$ and $e_{\mathrm{QSI}}$ rely on computing the dispersion of the ground state energy with $\bm{k}$ and $\bm{\phi}$, respectively.

With access to the ED spectra, we then use the fitting functions discussed in the main text to extract $e_{\mathrm{QSI}}$ and $c_{\mathrm{QSI}}$, as shown in Fig.~\ref{fig:e2_and_c_data}a and b in the main text. 
In Fig.~\ref{fig:e2_and_c_data}a, $e_{\mathrm{QSI}}$ is the only fitting parameter in the fitting function, equation~\ref{eqn:uE}. For all values of $(\mu,\zeta)$, we fit $e_\mathrm{QSI}$ using the ED data for $|Q\phi|^{2}<2$, where the electric field is small and thus linear electromagnetism applies. 
Our fit at $\mu=\zeta=0$, shown in the inset of Fig.~\ref{fig:e2_and_c_data}a, yields $e^{2}_{\mathrm{QSI}} = .04$ and has a covariance of $6\times 10^{-8}$. In fitting $c_{\mathrm{QSI}}$, we fit to the ED data for $|\bm{k}|a<\pi$ to avoid large momentum ED data with magnetic monopole states, which the dispersion function equation~\ref{eqn:dispersion} does not include. For $\mu=\zeta=0$, as shown in the inset of Fig.~\ref{fig:e2_and_c_data}b, we fit $c_{\mathrm{QSI}} = 0.51$ with a covariance of $1.7\times 10^{-5}$.

\textbf{Vacuum Sector Electric Field Energy} In the main text, we measure $e_{\mathrm{QSI}}$ by fitting Maxwell's electrodynamics to the ground state energy density as a function of the electric topological sectors $\bm{\phi} = \left(\phi_{1}, \phi_{2}, \phi_{3}\right)$. Here we show the relation between $\bm{\phi}$ and the corresponding electric field, $\bm{E}$. From this, we find the electric energy density, as given by Eqn.~\eqref{eqn:uE} in the main text.

The electric field component affiliated with, for example, $\phi_{1}$ is threaded through the parallelepiped's side spanned by $\bm{w_{2}}$ and $\bm{w_{3}}$. At a coarse-grained level, the periodic boundary conditions gives rise to translational invariance that allows us to use Gauss's law to solve for the corresponding uniform electric field. For each of the three independent sides of the parallelepiped, we have an expression relating the electric flux to the charge $e_{\mathrm{QSI}}$:
\begin{align}
\bm{E}\cdot(\bm{w_{1}}\cross\bm{w_{2}}) &= 4\pi\phi_{3} e_{\mathrm{QSI}},\\
\bm{E}\cdot(\bm{w_{2}}\cross\bm{w_{3}}) &= 4\pi\phi_{1} e_{\mathrm{QSI}},\\
\bm{E}\cdot(\bm{w_{3}}\cross\bm{w_{1}}) &= 4\pi\phi_{2} e_{\mathrm{QSI}}.
\end{align}
Solving for the electric field by introducing the volume of the parallelepiped $V = |\bm{w_{1}}\cdot(\bm{w_{2}}\cross\bm{w_{3}})|$, we find that
\begin{equation}
\bm{E} = 4\pi e_{\mathrm{QSI}}\frac{\phi_{1}\bm{w_{1}}+\phi_{2}\bm{w_{2}}+\phi_{3}\bm{w_{3}}}{V}.
\end{equation}
Let us now define a matrix, $Q$, whose column vectors are proportional to the vectors $\bm{w_{1}}$, $\bm{w_{2}}$, and $\bm{w_{3}}$,
\begin{equation}
Q \equiv \frac{a^{2
}}{V}
\begin{pmatrix}
\vert & \vert & \vert \\
 \bm{w_1}   & \bm{w_2} & \bm{w_3}   \\
 \vert & \vert & \vert
\end{pmatrix},
\end{equation}
where $a$ is the FCC lattice constant. $Q$ is a $3\times 3$ dimensionless matrix that incorporates the details of the parallelepiped's shape and size. Using this definition, we rewrite the electric field as
\begin{equation}
\bm{E} = 4\pi e_{\mathrm{QSI}}\frac{Q\bm{\phi}}{a^{2}}.
\end{equation}
Plugging this into Eqn.~\eqref{eqn:Maxwell} with $|\bm{B}| = 0$ gives us an expression for the electric field energy density.
\begin{equation}
    u = 2\pi\frac{|Q\bm{\phi}|^{2}}{a^{4}}e_{\mathrm{QSI}}^{2},
\end{equation}
This is the exact expression given by Eqn.~\eqref{eqn:uE} and is the fitting function shown in the inset of Fig.~\ref{fig:e2_and_c_data}a.\\

\textbf{Gaussian Photon Dispersion} 
The low-energy effective theory that describes the emergent photon in pyrochlore QSI is given by~\cite{Moessner2003, Hermele2004, Benton2012}
\begin{equation}\label{eqn:QED_RK}
    H = \frac{U}{2}\sum_{\bm{x},\mu} E_{\bm{x},\mu}^{2} +\frac{K}{2} \sum_{\hexagon}(\operatorname{curl}_{\hexagon} A)^{2} + \frac{V}{2}\sum_{\hexagon} (\operatorname{curl}_{\hexagon} E)^{2},
\end{equation}
where $(\bm{x},\mu)$ denotes an edge on the diamond lattice (the premedial lattice of the pyrochlore lattice) and $\operatorname{curl}_{\hexagon}$ is a lattice curl about a ``hexagonal'' plaquette on the diamond lattice, defined as the oriented sum of the link variable around a given hexagon on the diamond lattice. 
$E_{\bm{x},\mu}$ and $A_{\bm{x}, \mu}$ are the canonically conjugate electric field and vector potential operators, respectively. The electric field is related to the $S^{(z)}$ operator on the pyrochlore lattice by $E_{\bm{x},\mu} = \pm S^{(z)}_{\bm{x},\mu}$, where the plus or minus depends on whether the site $\bm{x}$ is an A or B site. the vector potential is the phase operator of $S^{\pm}$ and given by $S_{\bm{x},\mu}^{\pm} = e^{i(\pm)(\pm)A_{\bm{x},\mu}}$, where the first $\pm$ comes from the $\pm$ in $S^{(\pm)}$ and the second depends on whether the site $\bm{x}$ is an A or B site \cite{Hermele2004, Gingras2014}. The ice rule is translated to the divergence-free constraint $\operatorname{div}_{\bm{x}} E = 0$. Additionally, we note that from perturbation theory about the RK point, the first term near the RK point goes like $U \sim 1 - \mu$~\cite{Moessner2003}, ultimately vanishing at the RK point.

We measure the emergent speed of light using the photon dispersion calculated from the Eqn.~\eqref{eqn:QED_RK}. For completeness, here we follow Ref.~\cite{Benton2012} to re-derive the dispersion, as given by Eqn.~\eqref{eqn:dispersion} in the main text, using our notation and terminology of what excitations emit the emergent electric and magnetic fields. We refer the reader to Refs.~\cite{Hermele2004, Benton2012,Gingras2014, Szabo2019a,Kwasigroch2017} for a more thorough and detailed treatment of the calculation and corresponding discussion regarding the emergent gauge field and its dual formulation.

The electric field and vector potential operators $E_{\bm{x},\mu}$ and $A_{\bm{x},\mu}$ act on the edge of the diamond lattice, which we label with the notation $(\bm{x},\mu)$ where $\bm{x}$ is a site on the diamond lattice and $\mu$ the direction corresponding to the edge. Taking advantage of electromagnetic duality, the photon dispersion is calculated on the dual lattice, whose edges we denote $(\bm{y},\nu)$. Every edge (hexagonal plaquette) on the diamond lattice corresponds to a hexagonal plaquette (edge) on its dual. Rewriting Eqn.~\eqref{eqn:QED_RK} on the dual lattice gives
\begin{equation}\label{eqn:QED_RK_dn}
    H = \frac{U}{2}\sum_{\hexagon_{d}} E_{\hexagon_{d}}^{2} +\frac{K}{2} \sum_{\bm{y},\nu}(\operatorname{curl}_{\bm{y},\nu} A)^{2} + \frac{V}{2}\sum_{\bm{y},\nu} (\operatorname{curl}_{\bm{y},\nu} E)^{2},
\end{equation}
where the notation $\hexagon_{d}$ refers to hexagonal plaquettes on the dual lattice.

In the vacuum sector, where there are no electric charges or magnetic monopoles, there is an exact duality between the electric field operator on the diamond lattice, $E_{\hexagon_{d}}$, and the magnetic field operator on the dual lattice, $B_{\bm{y},\nu}$, which is defined in the typical way
\begin{equation}\label{eqn:dualB}
    B_{\bm{y},\nu} = \operatorname{curl}_{\bm{y},\nu}A.
\end{equation}
As discussed in the main text, the ice rule corresponds to a divergence free constraint $\operatorname{div}_{\bm{x}}E = 0$. This allows us to rewrite $E_{\hexagon_{d}}$ as
\begin{equation}\label{eqn:dualG}
    E_{\hexagon_{d}} = \operatorname{curl}_{\hexagon_{d}}G,
\end{equation}
where $G_{\bm{y},\nu}$ is the electric vector potential acting on the dual lattice edge $(\bm{y},\nu)$. It is dual to $A_{\hexagon_{d}}$ on the diamond lattice.

The operators $B_{\bm{y},\nu}$ and $G_{\bm{y},\nu}$ on the dual lattice play the role of $E_{\bm{x},\mu}$ and $A_{\bm{x},\mu}$ on the diamond lattice. Plugging in Eqns.~\eqref{eqn:dualB} and \eqref{eqn:dualG} in to Eqn.~\eqref{eqn:QED_RK_dn} gives
\begin{equation}\label{eqn:dualEffThy}
    \begin{aligned}
    H = &\frac{U}{2}\sum_{\hexagon_{d}} (\operatorname{curl}_{\hexagon_{d}}G)^{2} +\frac{K}{2} \sum_{\bm{y},\nu}B_{\bm{y},\nu}^{2}\\ 
        &\hspace{60pt}+ \frac{V}{2}\sum_{\bm{y},\nu} (\operatorname{curl}_{\bm{y},\nu} \operatorname{curl}_{\hexagon_{d}}G)^{2}.
    \end{aligned}
\end{equation}

We diagonalize the above Gaussian theory by introducing the photon creation (annihilation) operator $b^{\dagger}_{\bm{k},s}$ ($b_{\bm{k},s}$) which creates (destroys) a photon with momentum $\bm{k}$ and polarization $s$. They obey the usual Bose commutation relation $[b_{\bm{k},s},b_{\bm{q},s'}^{\dagger}] = \delta_{\bm{k},\bm{q}}\delta_{s,s'}$. 
Working in the Coulomb gauge, we write
\begin{align}
    \label{eqn:G}
    G_{\bm{y},\nu} &= \sqrt{\frac{4}{N}}\sum_{\bm{k},s} \sqrt{\frac{\hbar K}{2\omega_{\bm{k}s}}}\left[(\xi_{\bm{k}})_{\nu s} b_{\bm{k}s} + (\xi_{-\bm{k}}^{*})_{s\nu} b_{-\bm{k}s}^{\dagger}\right] e^{-i\bm{k}\cdot\bm{r}},\\
    \label{eqn:B}
    B_{\bm{y},\nu} &= \sqrt{\frac{4}{N}}\sum_{\bm{k},s} i\sqrt{\frac{\hbar\omega_{\bm{k}s}}{2K}}\left[(\xi_{\bm{k}})_{\nu s} b_{\bm{k}s} - (\xi_{-\bm{k}}^{*})_{s\nu} b_{-\bm{k}s}^{\dagger}\right] e^{-i\bm{k}\cdot\bm{r}},
\end{align}
where $\bm{r} = \bm{y} + \bm{e}_{\nu}/2$, $\xi_{\bm{k}}$ is the photon polarization tensor, $\omega_{\bm{k},s}$ is the photon dispersion, $\bm{e}_{\nu}$ is a primitive lattice vector of the diamond lattice such that $\bm{r}$ corresponds to the midpoint of a dual lattice edge, and $N$ is the number of diamond lattice sites.

Plugging in Eqns.~\eqref{eqn:G} and \eqref{eqn:B} into Eqn.~\eqref{eqn:dualEffThy} gives~\cite{Benton2012}
\begin{equation}\label{eqn:photonH}
    \begin{aligned}
    H =& \hbar\sum_{\bm{k},s}\left(\frac{VK\lambda^{2}_{\bm{k}s}}{4\omega_{\bm{k}s}} +\frac{UK\lambda_{\bm{k}s}}{4\omega_{\bm{k}s}} + \frac{\omega_{\bm{k}s}}{4} \right)\bigg[b_{\bm{k}s}b^{\dagger}_{\bm{k}s}+b^{\dagger}_{\bm{k}s}b_{\bm{k}s}\bigg]\\ 
    & +\left(\frac{VK\lambda^{2}_{\bm{k}s}}{4\omega_{\bm{k}s}} + \frac{UK\lambda_{\bm{k}s}}{4\omega_{\bm{k}s}}- \frac{\omega_{\bm{k}s}}{4}\right)\bigg[b_{\bm{k}s}b_{-\bm{k}s} + b^{\dagger}_{\bm{k}s}b^{\dagger}_{-\bm{k}s}\bigg],
    \end{aligned}
\end{equation}
where
\begin{align}
    \lambda_{\bm{k}0} &= \lambda_{\bm{k}1} = \lambda(\bm{k}),\\
    \lambda_{\bm{k}2} &= \lambda_{\bm{k}3} = 0,
\end{align}
with $\lambda(\bm{k})$ given in the main text after Eqn. \eqref{eqn:dispersion}. We find the dispersion $\omega_{\bm{k}s}$ such that the non-photon conserving term vanishes:
\begin{equation}
    \frac{WU\lambda^{2}_{\bm{k}s}}{4\omega_{\bm{k}s}} + \frac{KU\lambda_{\bm{k}s}}{4\omega_{\bm{k}s}}- \frac{\omega_{\bm{k}s}}{4} \equiv 0.
\end{equation}
Doing so, Eqn.~\eqref{eqn:photonH} can be rewritten as
\begin{equation}
    H = \sum_{\bm{k},s} \hbar\omega_{\bm{k}s} (b^{\dagger}_{\bm{k}s}b_{\bm{k}s} + 1/2). 
\end{equation}
The dispersion of the four polarization are given by
\begin{align}
    \omega_{\bm{k}0} &= \omega_{\bm{k}1} = \omega(\bm{k}),\\
    \omega_{\bm{k}2} &= \omega_{\bm{k}3} = 0,
\end{align}
where
\begin{equation}\label{eqn:apDispersion}
    \omega(\bm{k}) = \sqrt{UK\lambda(\bm{k}) + VK\lambda^{2}(\bm{k})}.
\end{equation}
Expanding the above for small $|k|$ gives $\omega(\bm{k}) = \sqrt{UK}\bm{k}a + \mathcal{O}(k^{2})$, which allows us to identify the emergent speed of light as $c_{\mathrm{QSI}} = a\sqrt{UK}$. Rewriting Eqn.~\eqref{eqn:apDispersion} in terms of $c_{\mathrm{QSI}}$ as well as defining $M\equiv VK$ yields the dispersion referenced by Eqn.~\eqref{eqn:dispersion} in the main text.

While fitting this dispersion to our ED data, we assume that $M$ is analytic in $\mu$ and $\zeta$ and approximate it by expanding to first order: $M = M_{0} + M_{\mu 1}\mu + M_{\zeta 1}\zeta$. We find that $M_{0} = -0.00667(1) $, $M_{\mu 1} = 0.050722(8)$, and $M_{\zeta 1} = -0.0606(1)$, where the interval in parenthesis gives the variance of the parameter fit. This is done to overcome challenges of measuring a vanishing speed of light at the RK point due to the lack of data at small $|\bm{k}|a$. In this fit, we find the $M_{\mu 1}$ coefficient by fitting the ED data with $c_\mathrm{QSI} = 0$ in the Eqn.~\eqref{eqn:dispersion}. $M_{0}$ is found doing a two parameter fit at $\zeta = \mu = 0$, and $M_{\zeta 1}$ with a two parameter fit at $\zeta = 1$.

\begin{figure*}[t!]
\centering
    \includegraphics[width=\textwidth]{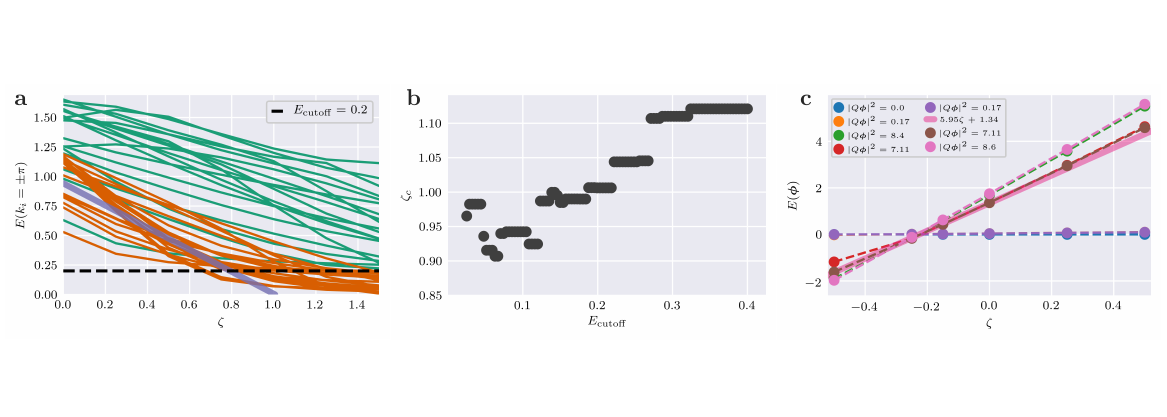}
    \caption{\textbf{a)} Following our extrapolation scheme discussed in the supplemental material, this shows the energy levels at $E_{\mathrm{cutoff}} = 0.2$ used to calculate $\zeta_{c}$. The lines in orange are those that have energy less than $E_{\mathrm{cutoff}}$ at $\zeta = 1.5$, and those in green is greater than this threshold energy. The purple line is the linear fit used in our extrapolation scheme to estimate $\zeta_{c}$, which is $\zeta_{c}=1$ in this case.
    \textbf{b)} Shows the different values of $\zeta_{c}$ at different threshold energies $E_{\mathrm{cutoff}}$. Averaging these different values of $\zeta$ yield the results that $\zeta_{c}\sim 1$ used in the main text. \textbf{c)} Shows the lowest energies of different electric topological sectors. We see that at $\zeta\sim -0.2$ a state with finite $|Q\bm{\phi}|$ becomes the new ground state, showing a phase transition to finite electric flux state at $\zeta \sim -0.2$.}
    \label{fig:zetaLevelCrossing}
\end{figure*}

\textbf{Extent of the deconfined phase} Previous work~\cite{Hermele2004, Shannon2012, Gingras2014} has established that the unperturbed QSI model ($\mu = \zeta = 0$), Eq.~\eqref{eqn:QSIeff}, lies in the deconfined eQED phase.  
It is further well established that this phase extends along the $\zeta = 0$ axis up to the RK point at $\mu = 1$ and down to $\mu \approx -0.5$~\cite{Shannon2012}, outside of which the system transitions into confining ordered states. 
Our numerical investigations are consistent with these expectations, finding evidence of a first order transition into a different topological sector at $\mu \approx -0.5$ and the RK transition at $\mu = 1$.

The two-body $\zeta$ term has been briefly discussed previously in the literature \cite{Savary2012, Chen2016}, but has yet to be numerically investigated. 
In particular, the extent of the eQED phase along this axis has not been previously studied. 
Here we elaborate on the claim that Eqn.~\eqref{eqn:QSIeff} with $\mu = 0$ in Eqn.~\eqref{eqn:QSIp} realizes an eQED phase for $-0.2\lesssim \zeta \lesssim 1$. 

As $\zeta$ is increased, ED data in the $\bm{k} = (\pm \pi, \pm\pi,\pm\pi)$ momentum sector (the L point of the Brillouin zone) approach zero. 
These points correspond to the clear vertical line of data in the inset of Fig.~\ref{fig:e2_and_c_data}b at $|\bm{k}|a = \sqrt{3}\pi$. 
This suggests a phase transition into an ordered state at finite momentum. 
We expect the new ground state to possess long-range antiferromagnetic order and, in consideration of the excitation spectra at finite $|\bm{k}|$, that the transition may be understood in terms of magnetic monopole condensation from the eQED phase. 
This is consistent with theoretical work that found the condensation of magnetic monopoles in a pure $U(1)$ gauge theory on the diamond lattice corresponds to the development of antiferromagnetic order in QSI, but with $|\bm{k}|a = 2\pi$ ~\cite{Chen2016}.

We estimate the value of $\zeta$ at which this transition occurs, $\zeta_{c}$, by an extrapolation scheme. 
First, consider the lowest energy states within the $\bm{k} = (\pm \pi, \pm\pi,\pm\pi)$ sector.
The energy of these states decrease linearly as $\zeta$ increases and then saturate toward 0, signifying the phase transition at some $\zeta_{c}$.
We first pick a value of $\zeta>\zeta_{c}$ at which the antiferromagnetic ground state is clearly developed.
We chose $\zeta = 1.5$, but other values of $\zeta$ could be equivalently chosen without significantly changing our conclusion.
We then pick a threshold energy, $E_{\mathrm{cutoff}}$, and consider only the states whose energy at $\zeta = 1.5$ is less than $E_{\mathrm{cutoff}}$. 
The introduction of $E_{\mathrm{cutoff}}$ is done to ensure that we only consider the energy levels that approach zero energy at the phase transition and therefore correspond to the new ground state with antiferromagnetic order. 
These energy levels are selected from our entire sample of shapes and follow similar qualitative behavior: they linearly decrease for increasing $\zeta$ and then for some $0<\zeta<\zeta_{c}$ saturate towards zero.
In Fig.~\ref{fig:zetaLevelCrossing}a, we show these energy levels for $E_{\mathrm{cutoff}}=0.2$.
Energy levels in orange correspond to states with $E<E_{\mathrm{cutoff}}$ at $\zeta = 1.5$.
Our extrapolation scheme works as follows.
Fitting a linear function ($A(\zeta + \zeta_{c})$) to the energies of these states for $\zeta < 0.5$, we extract the value of $\zeta_{c}$.
This fit is shown in the linear purple line in Fig.~\ref{fig:zetaLevelCrossing}a, and yields $\zeta_{c} = 1$.
We repeat this scheme at different $E_{\mathrm{cutoff}}$ values, calculating $\zeta_{c}$ as a function of $E_{\mathrm{cutoff}}$.
Fig.~\ref{fig:zetaLevelCrossing}b shows the different calculated values of $\zeta_{c}$ as a function of $E_{\mathrm{cutoff}}$ and suggests that the phase transition occurs somewhere between $0.9<\zeta<1.1$. 
From this, we extract the critical point  $\zeta_{c}\sim 1$. 
We note that a more systematic finite size scaling study of the transition is inaccessible to our methods and beyond the scope of this work.

We find that at $\zeta \sim -0.2$, there is a transition to a finite electric field state (ie. the ground state moves to a non-zero topological sector) and at $\zeta\sim 1$ a phase transition to a finite momentum state.
Fig.~\ref{fig:zetaLevelCrossing}c shows the lowest energy states within several of the electric topological sectors. 
At $\zeta\sim -0.2$, a nonzero $|Q\bm{\phi}|$ block becomes the new ground state, indicating that there is a phase transition from the eQED into an ordered phase. 
Within the projected ice manifold, this transition appears first order. We note that similar transitions to finite flux states occur both at $\mu \sim -0.5$ and $\mu = 1$. The phase transition at the RK point occurs along with the ground state energy in each topological sector becoming degenerate. This does not occur at $\mu\sim -0.5$ and $\zeta \sim -0.2$, leading us to believe the transition is first order.
Again, more systematic study of the phase and transition are beyond the scope of this work.

\end{document}